\documentclass[10pt,onecolumn,draftclsnofoot]{IEEEtran}

\usepackage[final]{graphicx}
\usepackage[reqno]{amsmath}
\usepackage{amssymb}
\usepackage{amsmath}
\usepackage{subfig}
\usepackage{epstopdf}
\usepackage{xcolor}
\usepackage{float}
\usepackage[paper=letterpaper,tmargin=1in,bmargin=.8in,left=.8in, right=.8in]{geometry}
\usepackage{comment}
\usepackage{ dsfont }
\usepackage{cite}

\begin{document}

\sloppy
 \pagestyle{empty}
\title{Fast Deep Learning for Automatic Modulation Classification}
\author{{\large{Sharan Ramjee, {\em Student Member, IEEE}, Shengtai Ju, {\em Student Member, IEEE}, \\Diyu Yang, {\em Student Member, IEEE}, Xiaoyu Liu, {\em Student Member, IEEE},\\Aly El Gamal, {\em Member, IEEE} and Yonina C. Eldar, {\em Fellow, IEEE}}}
\thanks{This work is supported in part by DARPA and AFRL under grant no. 108818, and was presented in part at the Asilomar Conference on Signals, Systems and Computers~\cite{asilomar17}.

S. Ramjee, S. Ju, D. Yang, X. Liu and A. El Gamal are with the Department of Electrical and Computer Engineering, Purdue University, West Lafayette, IN, 47907 USA (e-mail: sramjee, ju10, yang1467, liu1962, elgamala@purdue.edu).

Y. C. Eldar is with the Department of Electrical Engineering, Technion, Israel Institute of Technology, Haifa, Israel (e-mail: yonina@ee.technion.ac.il).}
}
\maketitle

\begin{abstract}
In this work, we investigate the feasibility and effectiveness of employing deep learning algorithms for automatic recognition of the modulation type of received wireless communication signals from subsampled data. Recent work considered a GNU radio-based data set that mimics the imperfections in a real wireless channel and uses 10 different modulation types. A Convolutional Neural Network (CNN) architecture was then developed and shown to achieve performance that exceeds that of expert-based approaches. Here, we continue this line of work and investigate deep neural network architectures that deliver high classification accuracy. We identify three architectures - namely, a Convolutional Long Short-term Deep Neural Network (CLDNN), a Long Short-Term Memory neural network (LSTM), and a deep Residual Network (ResNet) - that lead to typical classification accuracy values around 90\% at high SNR. We then study algorithms to reduce the training time by minimizing the size of the training data set, while incurring a minimal loss in classification accuracy. To this end, we demonstrate the performance of Principal Component Analysis in significantly reducing the training time, while maintaining good performance at low SNR. We also investigate subsampling techniques that further reduce the training time, and pave the way for online classification at high SNR. Finally, we identify representative SNR values for training each of the candidate architectures, and consequently, realize drastic reductions of the training time, with negligible loss in classification accuracy. 
\end{abstract}
\section{Introduction}\label{sec:intro}

Automatic modulation classification plays an important role in modern wireless communication. It finds applications in various commercial and military areas. For example, Software Defined Radios (SDR) use blind recognition of the modulation type to quickly adapt to various communication systems, without requiring control overhead. In military settings, friendly signals should be securely received, while hostile signals need to be efficiently recognized typically without prior information. Under such conditions, advanced real-time signal processing and blind modulation recognition techniques are required. Modulation recognition may also prove to be an essential capability for identifying the source(s) of received wireless signals, which can enable various intelligent decisions for a context-aware autonomous wireless communication system.

A typical modulation classifier consists of two steps: signal preprocessing and classification algorithms. Preprocessing tasks may include noise reduction and estimation of signal parameters such as carrier frequency and signal power. In the second step, three popular categories of modulation recognition algorithms are conventionally selected: likelihood-based (LB)\cite{sills1999maximum,polydoros1990detection,sapiano1996maximum,beidas1998asynchronous,panagiotou2000likelihood,hong2002antenna}, feature-based (FB)\cite{hsue1989automatic,hong1999identification,swami2000hierarchical,hatzichristos2001hierarchical,soliman1992signal,lichun2002comments} or using an artificial neural network (ANN)\cite{mingquan1998ar,mobasseri2000digital,mingquan1996cyclic,azzouz1996modulation,nolan2001modulation}. The first compares the likelihood ratio of each possible hypothesis against a threshold, which is derived from the probability density function of the observed wave. Multiple likelihood ratio test (LRT) algorithms have been proposed: Average LRT\cite{kim1988digital}, Generalized LRT\cite{lay1994per}, Hybrid LRT\cite{hong2002antenna} and quasi-hybrid LRT\cite{sills1999maximum}. In the FB approach, several features are selected and observed for the decision. Both LB and FB methods require precise estimates in the first step and have only been derived to distinguish between few modulation types\cite{kim1988digital,sapiano1996maximum,park2008automatic,de2010prototype}. ANN structures such as multi-layer perceptrons (MLP) have been widely used as modulation type classifiers\cite{mingquan1998ar}. Unlike LB and FB techniques, where the decision threshold is chosen manually, the threshold in ANN could be determined adaptively and automatically. Traditional MLP has performed well on modulation types such as AM, FM, ASK, and FSK. Recent work has shown that deep neural networks with cutting-edge structures could greatly improve the classification process (see e.g., \cite{conv} and \cite{resnet}).


Deep neural networks have played a significant role in the research domain of video, speech and image processing in the past few years. 
The recent success of deep learning algorithms is associated with applications that suffer from inaccuracies in existing mathematical models and enjoy the availability of large data sets. Recently, the idea of deep learning has been introduced for modulation classification using a Convolutional Neural Network (CNN)
for distinguishing between 10 different modulation types~\cite{conv}. Simulation results show that a CNN not only demonstrates better accuracy results, but also provides more flexibility in detecting various modulation types compared to current expert-based approaches. 
Residual Networks (ResNet) \cite{resnet} and Densely Connected Networks (DenseNet) \cite{densenet} were recently introduced to strengthen feature propagation in the deep neural network by creating shortcut paths between different layers in the network. By adding the bypass connections, an identity mapping is created, allowing the deep network to learn simple functions. A ResNet architecture was shown to be successful for distinguishing between 24 different modulation types in~\cite{new-resnet}. DenseNet performed well for image recognition, but has not been used in the area of modulation recognition. A Convolutional Long Short-term Deep Neural Network (CLDNN) has been recently introduced in~\cite{CLDNN}, combining the architectures of CNN and Long Short-Term Memory (LSTM) into a deep neural network by taking advantage of the complementarity of CNNs, LSTMs, and conventional deep neural network architectures. The LSTM unit is a memory unit of a Recurrent Neural Network (RNN). RNNs are neural networks with memory that are suitable for learning sequence tasks such as speech recognition and handwritten recognition. LSTM optimizes the gradient vanishing problem in RNNs by using a forget gate in its memory cell, which enables the learning of long-term dependencies. The authors in \cite{west2017deep} added LSTM units into the neural network model and presented high classification accuracy for a wide range of modulation formats. In this work, we present five different architectures that deliver higher classification accuracy than the CNN introduced in~\cite{conv}. We design our own CNN, DenseNet, and CLDNN architectures for the modulation recognition task, as well as derive optimized versions of the ResNet architecture of~\cite{new-resnet} and the LSTM architecture of~\cite{west2017deep}, by tuning the number of residual stacks for ResNet and the hyperparameters for LSTM.

One major challenge facing machine learning algorithms based on deep neural network architectures is the long training time. For example, for the problem at hand, even the simple CNN architecture in \cite{conv} would take approximately 40 minutes to train using three Nvidia Tesla P100 GPU chips. This creates a serious obstacle towards the feasibility of applying such algorithms in real-time, where online training is needed to adapt the network architecture to changing environmental conditions. In particular, applying deep learning to the autonomous wireless communication systems anticipated in next-generation networks require significant reductions in the training time compare to the state of the art methods. In such systems, it is likely that training of the machine learning algorithms will be frequently needed to accommodate new environmental conditions. Hence, the issue of reducing the training time becomes essential for the success of these algorithms. 

The objective of this work is two-fold: identifying deep neural network architectures suitable for the modulation classification task, and suggesting methods for reducing their training time. First, we study different deep neural network architectures for the task of modulation classification, based on the data set provided in \cite{datagen}. Inspired by recent studies, we explore five different architectures that deliver higher classification accuracy than the CNN architecture of~\cite{conv}, and build insights for their optimal designs as well as identify candidate architectures that deliver high classification accuracy for a wide range of SNR values. In particular, we find the CLDNN developed in this work and an optimized variant of the ResNet of~\cite{new-resnet} to deliver superior performance at low SNR. We also show that ResNet and an optimized variant of the LSTM architecture of~\cite{west2017deep} perform best at high SNR. We then explore various methods for reducing the training time by minimizing the size of the training set, while preserving relevant information to the classification task. These methods are based on dimensionality reduction and sub-Nyquist techniques (see \cite{yonina-book} for a review), as well as finding representative SNR values that are ideal for training. We derive insights for the impact and optimal designs of each of these methods. Our results confirm the possibility of reducing the training time by as much as 20 times, while incurring a minimal losses in classification accuracy (as low as 2\%). 

For the CLDNN, ResNet and LSTM architectures that are identified to perform best over different SNR ranges within the range from -20 dB to 18 dB, the training time drops linearly with the dimensionality reduction factor or the subsampling rate, as well as when reducing the number of example vectors in the training data sets through SNR selection. We find that reducing the input vector dimensions through Principal Component Analysis (PCA) is more effective than subsampling at low SNR, and the opposite holds at high SNR. In particular, we develop a subsampling method based on eliminating samples that have low magnitude values and show that this method leads to little degradation in classification accuracy at high SNR. Finally, we demonstrate the effectiveness of choosing representative training SNR values, and show that for the considered range of 20 SNR values from -20 dB to 18 dB, choosing a pair of SNR values can lead to benign accuracy degradation (less than 2\% for LSTM) with a 10 fold reduction in training time. The results suggest that a combination of two or more of these methods could be very powerful. For example, using uniform subsampling with a factor of 2, with training SNR values of 18 dB and 0 dB for the proposed LSTM architecture may lead to negligible accuracy degradation with a 20 fold reduction in training time. 

The rest of the paper is organized as follows. We first describe the considered data set and programming environment in Section~\ref{sec:setup}, and investigate deep neural network architectures and their classification performance in Section~\ref{sec:architectures}. We then consider the problem of minimizing the training time for three candidate architectures, while incurring minimal loss in accuracy. We start this study by investigating different methods to compress the input data through dimensionality reduction and subsampling in Section~\ref{sec:dimred}. We then explore in Section~\ref{sec:snr} the existence of representative SNR values, that could be used exclusively for training, while maintaining accurate classification over the whole tested SNR range. In Section~\ref{sec:discussion}, we discuss insights obtained from the presented results. We finally conclude this work in Section~\ref{sec:conclusion}.

\section{Experimental Setup}\label{sec:setup}
In this work, we consider the classification of the modulation type of received wireless signals, using deep neural network classifiers that adaptively incorporate features extracted from a training data set. A general expression for the received baseband complex envelope is 
\begin{equation}\label{eq1}
r\left(t\right)=s(t;{\boldsymbol u}_\mathbf i)+n\left(t\right),
\end{equation}
where
\begin{equation}\label{eq2}
s(t;{\boldsymbol u}_\mathbf i)\;=\;a_ie^{j2\pi\triangle ft}e^{j\theta}{\textstyle\sum_{k=1}^K}e^{j\phi_k}s_k^{(i)}g(t-(k-1)T-\varepsilon T),\;0\leq t\leq KT
\end{equation}
is the noise-free baseband complex envelope of the received signal, and $n(t)$ is the instantaneous channel noise at time $t$. In \eqref{eq2}, $a_i$ is the unknown signal amplitude, $\Delta f$ is the carrier frequency offset, $\theta$ is the time-invariant carrier phase introduced by the propagation delay, $\phi_k$ is the phase jitter, $\{s_k^{(i)}, 1\leq k \leq K\}$ denotes $K$ complex symbols taken from the $i^{th}$ modulation format, $T$ represents the symbol period, $\varepsilon$ is the normalized epoch for time offset between the transmitter and signal receiver, $g(t)\;=\;P_{pulse}(t)\otimes h(t)$ is the composite effect of the residual channel with $h(t)$ denoting the channel impulse response and $\otimes$ denoting mathematical convolution, and $P_{pulse}(t)$ is the transmit pulse shape. Here, ${\boldsymbol u}_\mathbf i\;=\;\{a_i,\;\Delta f,\;\theta,\;\varepsilon,\;g(t),\;\{\phi_k\}_{k=1}^{K},\;\{s_k^{(i)}\}_{k=1}^K\}$ is the multidimensional vector that includes the deterministic unknown signal or channel parameters for the $i^\textrm{th}$ modulation type. \textbf{The goal is to recognize the modulation type $i$ from the received signal $r(t)$}. To this end, we use various machine learning classifiers based on deep neural network architectures, where a training data set is first processed to set the network parameters, and then the classification accuracy is computed over the classification output for a testing data set (see~\cite{dl-book} for more details).

\begin{figure}
\centering
	\includegraphics[width=5in]{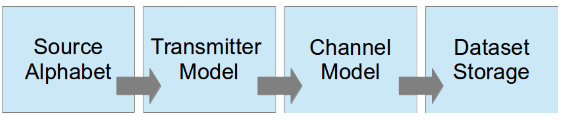}
	\caption{A frame of data generation.}
	\label{fig:frame}
\end{figure}

We use the RadioML2016.10b data set generated in~\cite{conv} as the input data of our research. Details about the generation of this data set can be found in \cite{datagen}. Fig. \ref{fig:frame} shows a high-level framework of the data generation. For digital modulations, the entire Gutenberg works of Shakespeare in ASCII is used, with whitening randomizers applied to ensure equiprobable symbols and bits. For analog modulations, a continuous voice signal is used as input data, which consists primarily of acoustic voice speech with some interludes and off times. Ten widely used modulations are chosen: eight digital and two analog modulations. These consist of BPSK, QPSK, 8PSK, QAM16, QAM64, BFSK, CPFSK, and PAM4 for digital modulations, and WB-FM, and AM-DSB for analog modulations. The data set is split equally among all considered modulation types. For the channel model, physical environmental noises like thermal noise and multipath fading were simulated. The models for generating random channel and device imperfections include sample rate offset model, noise model, center frequency offset model and fading model. When packaging data, the output stream of each simulation is randomly segmented into vectors as the original data set with a sample rate of 1M sample per second. Similar to the way that an acoustic signal is windowed in voice recognition tasks, a sliding window extracts 128 samples with a shift of 64 samples, which forms the data set we are using. 160,000 samples generated using the GNU-radio library developed in~\cite{datagen} are segmented into training and testing data sets through 128-samples rectangular windowing processing, which is similar to the  windowed continuous acoustic voice signal used in voice recognition tasks. The training examples - each consisting of 128 samples - are fed into the neural network in 2$\times$128 vectors with real and imaginary parts separated in complex time samples, except for the pure LSTM architecture, which is fed samples in polar form (amplitude and phase). The labels in input data include the SNR ground truth and the modulation type. The SNR of the samples is uniformly distributed from -20dB to +18dB, with a step size of 2 dB, i.e., the data set is equally split among all SNR dB values in $\{-20,-18,-16,\cdots,18\}$. Finally, the classification accuracy is measured as the percentage of correctly classified samples over the testing data set.

For all our experiments, we used Keras with TensorFlow as a backend\footnote{Code is available at: https://github.com/dl4amc/source}. We used a GPU server equipped with 3 Tesla P100 GPUs with 16 GB memory. The Adam optimizer was used for all architectures, and the loss function was the categorical cross entropy function. We also used ReLu activation functions for all layers, except the last dense layer, where we used Softmax activation functions. For all architectures except the LSTM, we used a batch size of 1024, and a learning rate of 0.001. For the LSTM architecture, we used a batch size of 400, and a learning rate of 0.0018. 



\section{Deep Neural Network Architectures}\label{sec:architectures}
We investigated the performance of five different types of neural network architectures on the considered problem of modulation classification: Convolutional Neural Network (CNN), Densely Connected Convolutional Network (DenseNet), Convolutional Long Short-term Memory Deep Neural Network (CLDNN), Long Short-term Memory Network (LSTM), and Deep Residual Network (ResNet).

\subsection{CNN and DenseNet Architectures}\label{sec:cnn}
\begin{figure}[!htb]
   \begin{minipage}{0.48\textwidth}
     \centering
     \includegraphics[width=\linewidth]{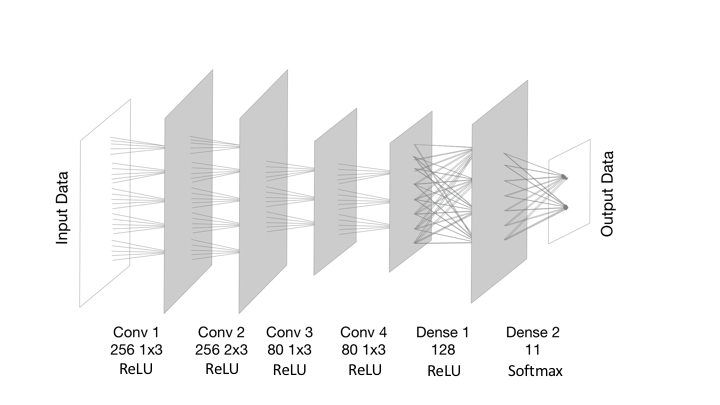}
     \caption{CNN architecture.}\label{fig:cnn4}
   \end{minipage}\hfill
   \begin{minipage}{0.48\textwidth}
     \centering
     \includegraphics[width=\linewidth]{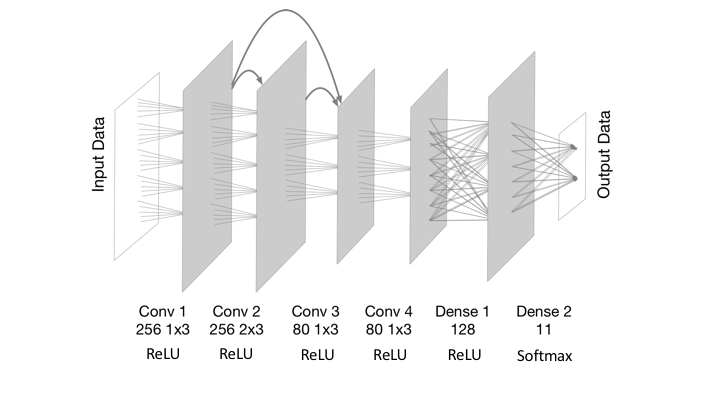}
     \caption{DenseNet architecture.}\label{fig:densenet}
   \end{minipage}
\end{figure}

We start with a convolutional neural network architecture similar to the CNN2 network proposed in \cite{conv}, which achieves an accuracy of 75\% at high SNR. We experiment with different network depths and filter settings. The best accuracy we achieved at high SNR was approximately 83.8\%, using a CNN architecture with four convolutional layers, as shown in Fig.~\ref{fig:cnn4}. The first parameter below each convolutional layer in the figure represents the number of filters in that layer, while the second and third numbers show the size of each filter. For the two dense layers, there are 128 and 11 neurons, in order of their depth in the network. We note that the improved performance is due to the extra two convolutional layers, compared to CNN2.

We next investigate the effect of shortcut connections by introducing a DenseNet architecture shown in Fig.~\ref{fig:densenet}. The architecture of DenseNet is similar to that of a CNN, except for the shortcut connections between non-consecutive layers. This architecture achieves an improved classification accuracy of 86.6\% at high SNR.


\subsection{CLDNN Architecture}\label{sec:cldnn}
\begin{figure}[!htb]
   \begin{minipage}{0.48\textwidth}
     \centering
     \includegraphics[width=\linewidth]{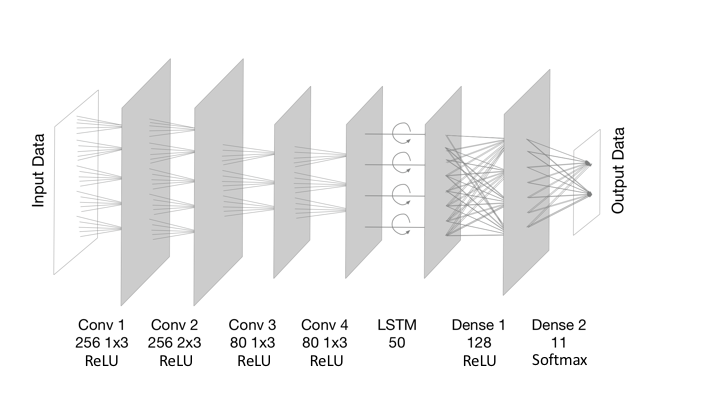}
     \caption{CLDNN architecture.}\label{fig:cldnn}
   \end{minipage}\hfill
   \begin{minipage}{0.48\textwidth}
     \centering
     \includegraphics[width=\linewidth]{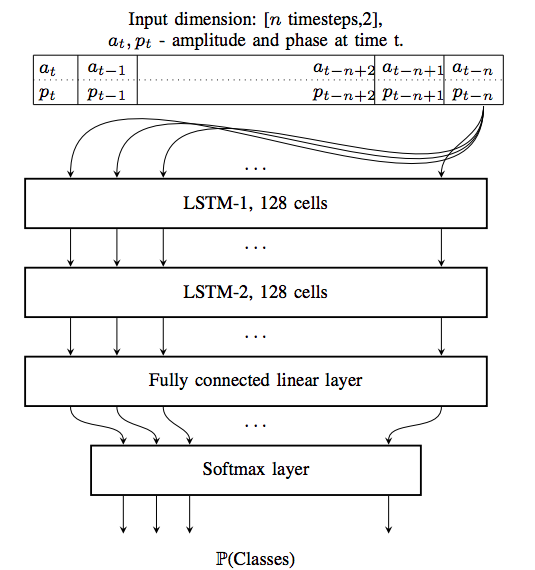}
     \caption{LSTM architecture.}\label{fig:lstm}
   \end{minipage}
\end{figure}

Recurrent Neural Networks (RNN) have been proved to provide a powerful tool for time domain data processing tasks due to their ability to connect previous state information to the present task. Inspired by \cite{CLDNN}, we propose a CLDNN architecture by adding an LSTM layer into the CNN architecture. The detailed architecture considered for CLDNN is shown in Fig. \ref{fig:cldnn}. The extra LSTM layer is placed between the CNN layers and the dense layers. In our experiments, an LSTM layer with 50 cells provided the best accuracy. This architecture results in a classification accuracy of 88.5\% at higher SNR values (above 2 dB). The RNN structure is suited for modulation classification, because it can extract temporal relationships from the input waveform.

\subsection{LSTM Architecture}\label{sec:lstm}
In \cite{lstm}, the authors proposed a modulation classification model based on a pure LSTM architecture. The design of this network is based on similar intuition as our CLDNN, namely that LSTM is efficient in learning long-term dependencies in time series data processing tasks. However, unlike CLDNN, LSTM does not have convolutional layers. This architecture \textbf{receives the input samples in polar form}, instead of the rectangular form used for all other considered architectures. The polar form representation is obtained by computing the amplitude and phase of the input I/Q sample at each sampling time step. It then uses two LSTM layers, each with 128 cells, to extract the temporal dependencies of the amplitude and phase characteristics of different modulation schemes. It uses a dense layer with Softmax activation function as the last hidden layer to project the output of the second LSTM layer into the final probability output space $\mathds{P(\text{Classes})}$.

We fine tuned the hyperparameters of the LSTM network in \cite{lstm} and found out that it achieves a classification accuracy of 92\% at high SNR. The great performance of the LSTM network further demonstrates that RNNs provide a good choice for the task of modulation classification in terms of classification accuracy, due to their ability to extracting long-term temporal relationships in time series data, which could be useful for identifying patterns of symbol-to-symbol transitions. 

However, RNNs suffer from several issues when it comes to online learning. 
First, the training time for RNNs is much slower than the training time of Feed-forward Neural Networks (FNNs). The training time for the LSTM network is 222 seconds per epoch using all 3 GPUs, about 4 times longer than the training time of the CNN network with four convolutional layers. The reason is two-fold: First, the computational complexity of the optimization process of RNNs is much greater than that of traditional FNNs. RNNs perform one full optimization step (a forward propagation and a backpropagation) for each time step of each batch, while FNNs only perform one full optimization step per batch. Second, the computations of FNNs can usually be easily parallelized, while it is harder to do so for computations of RNNs, since each time domain step of RNN computation depends on previous steps. Such long training time becomes a bottleneck in online training, where data comes in real time and the training process needs to be done fast.
Second, the temporal relationships that RNNs attempt to extract are likely to break down either when the sampling rate of the arriving signals is not fixed, or when a dimensionality reduction or subsampling process is necessary for achieving faster learning. This is demonstrated in more details in Section \ref{sec:dimred}, where we see that the classification accuracy for both CLDNN and LSTM networks drops rapidly at high SNR when performing subsampling or dimensionality reduction. In these cases, the ResNet architecture presented below performs best.

\subsection{ResNet Architecture}\label{sec:resnet}
As neural networks grow deeper, their learning performance has been challenged with problems like vanishing or exploding gradient and overfitting, and therefore both training and testing accuracy for a deep neural network start to degrade after the network reaches a certain depth. The degradation of testing accuracy results from the overfitting issue, due to the extra complexity in a deep neural network, and the degradation of training accuracy is due to the problem of vanishing/exploding gradient that makes the optimizer less viable to converge to a sufficiently good local minimum of the cost function.

\begin{figure}[!htb]
   \begin{minipage}{0.48\textwidth}
     \centering
     \includegraphics[width=\linewidth]{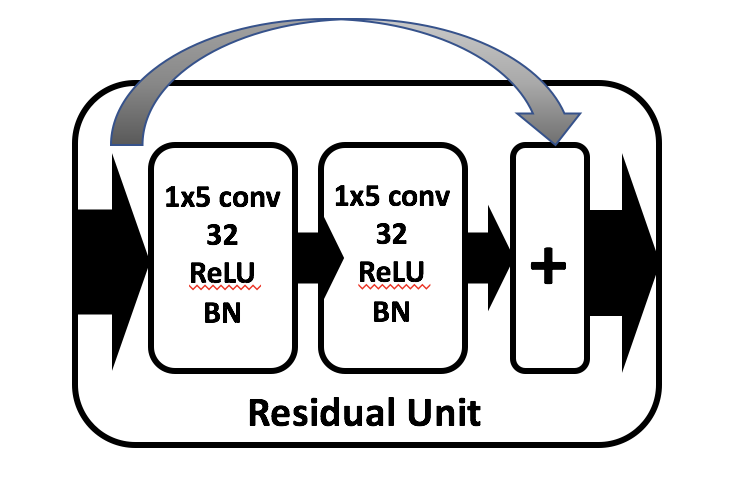}
     \caption{Residual unit architecture.}\label{fig:res_unit}
   \end{minipage}\hfill
   \begin{minipage}{0.48\textwidth}
     \centering
     \includegraphics[width=\linewidth]{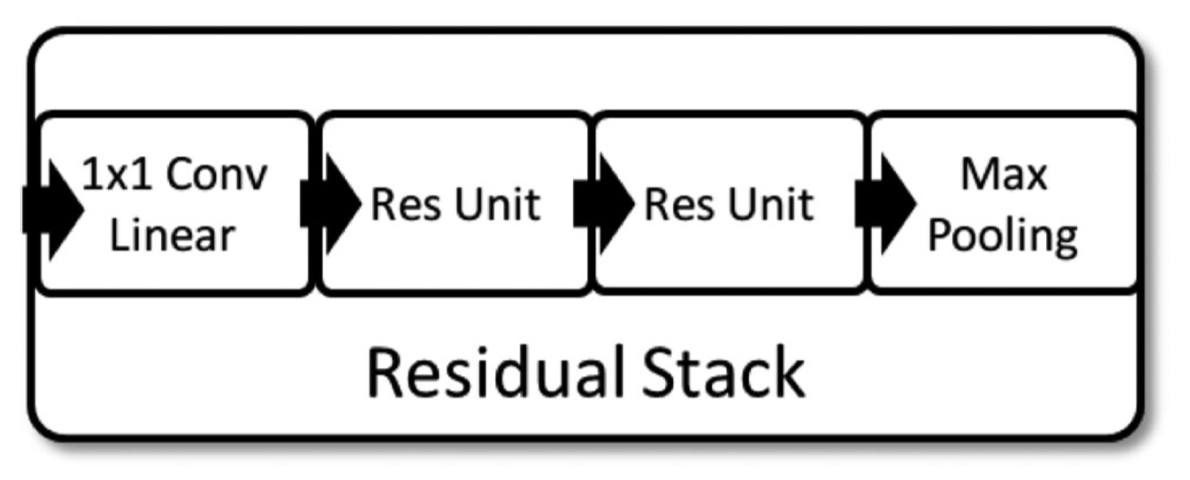}
     \caption{Residual stack architecture.}\label{fig:res_stack}
   \end{minipage}
\end{figure}

The Deep residual Network (ResNet) architecture was introduced in ImageNet and COCO 2015 competitions\cite{resnet}. It resolves accuracy degradation issues in deeper neural networks and has been shown to be a robust choice for a wide range of machine learning tasks. 
Inspired by the ResNet architecture in \cite{new-resnet}, we design a similar ResNet, but with three residual stacks instead of six, as we found that choice to lead to increasing the classification accuracy. The overall architecture is shown in Table \ref{table:resnet}. In our network, three residual stacks are followed by three fully connected layers, where each residual stack consists of one convolutional layer, two residual units, and one max-pooling layer. For each residual unit, a shortcut connection is created by adding the input of the residual unit with the output of the second convolutional layer of the residual unit. Each convolutional layer in the residual unit uses a filter size of 1x5 and is followed by a batch normalization layer to prevent overfitting. The detailed structures of residual units and residual stacks are shown in Fig. \ref{fig:res_unit} and Fig. \ref{fig:res_stack}, respectively. Compared to the DenseNet architecture in Section \ref{sec:cnn}, the proposed ResNet has reduced complexity in each layer. Therefore, ResNet is able to go deeper without experiencing accuracy degradation issues. The ResNet architecture delivers a classification accuracy of 92\% at high SNR with a fast training speed of 58 seconds per epoch.

\begin{figure}
	\includegraphics[width=\linewidth]{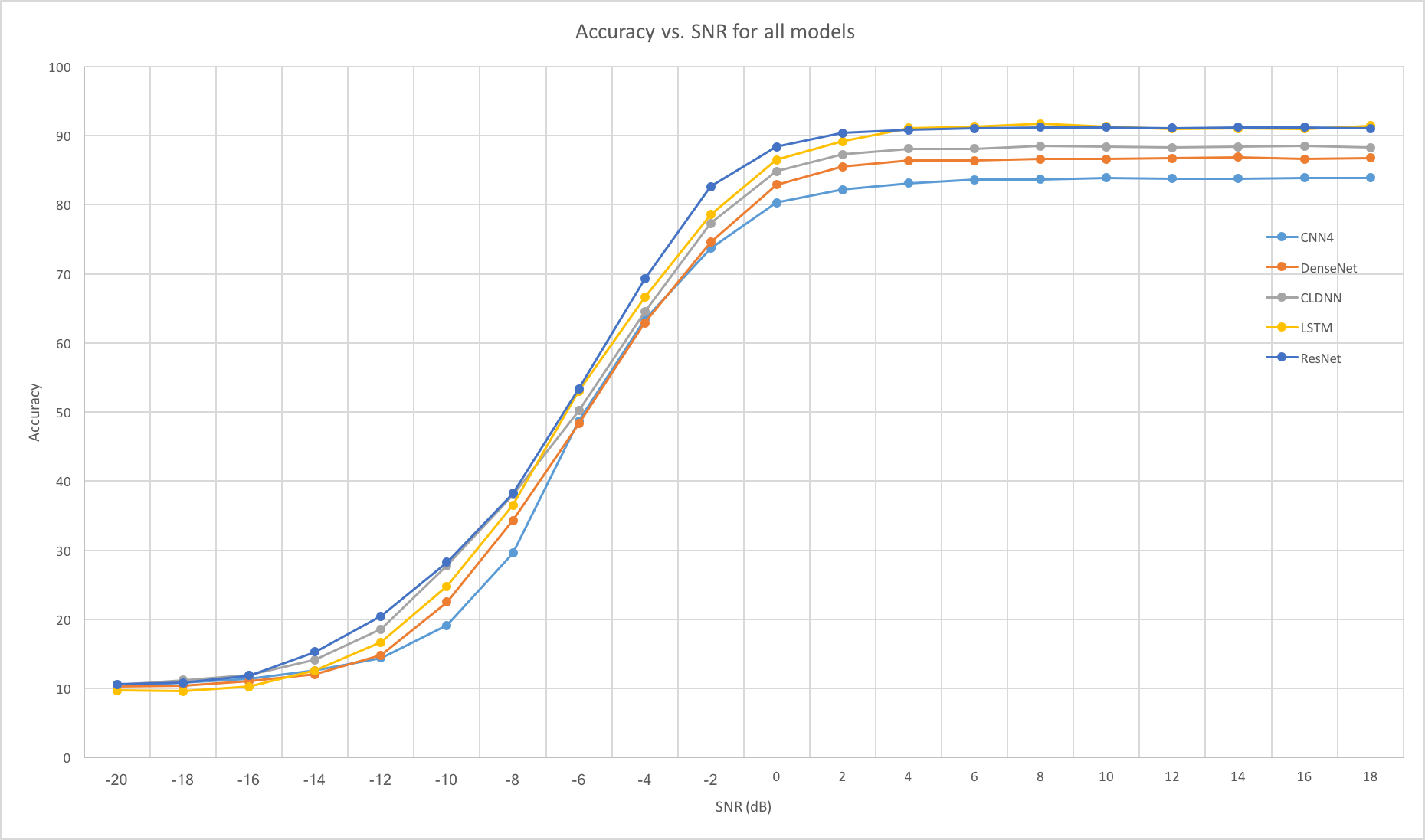}
	\caption{Accuracy vs SNR all models.}
	\label{fig:accuracy_all}
\end{figure}

We show the overall accuracy versus SNR results for all models in Fig. \ref{fig:accuracy_all}. We identify three architectures that deliver good performance: CLDNN, LSTM, and ResNet. At high SNR, both LSTM and ResNet achieve similar classification accuracies of 92\%, which is the best result among all models. At low SNR, CLDNN, and ResNet deliver the best results. 

\begin{table}[t]
\begin{center}
 \begin{tabular}{||c c||} 
 \hline
 Layer & Output dimensions \\ [0.5ex] 
 \hline\hline
 Input & 2x128 \\ 
 Residual Stack & 32x64\\
 Residual Stack & 32x32\\
 Residual Stack & 32x16\\
 FC/ReLU & 128\\
 FC/ReLU & 128\\
 FC/Softmax & 10\\[1ex]
 \hline
\end{tabular}
\end{center}
\caption{ResNet Architecture.}
\label{table:resnet}
\end{table}

\section{Dimensionality Reduction and Subsampling}\label{sec:dimred}
\begin{figure}[htb]
	\includegraphics[width=\linewidth]{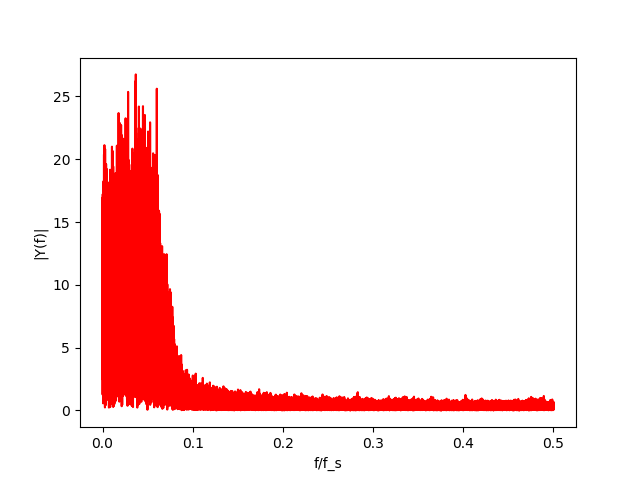}
	\caption{One-sided normalized FFT for a BPSK signal. A value of $0.5$ on the horizontal axis corresponds to the Nyquist rate (Bandwidth is half the sampling rate). Most of the signal energy is within a band of around $\frac{1}{12}$ of the sampling rate.}
	\label{fig:fft}
\end{figure}

In this section, we present various attempts to minimize the training time, by reducing the dimensionality of each vector sample input to the deep neural network classifier. Based on the results in Section~\ref{sec:architectures}, we consider the CLDNN, ResNet and LSTM architectures described in Sections~\ref{sec:cldnn}, \ref{sec:resnet} and \ref{sec:lstm}, respectively. For each considered method for minimizing the training time, we report the results obtained by reducing the number of input vector dimensions by factors of $2^k, 1 \leq k \leq 5$. Recall that each training example input vector has originally 256 dimensions, with each of the 128 complex time samples occupying 2 dimensions. It is important to note from the frequency domain representation of the input waveform depicted in Fig. \ref{fig:fft} that the sampling rate of the input waveform is around 6 times the Nyquist rate, and hence, the reduction goes to sub-Nyquist levels only for factors of 8, 16 and 32.

\subsection{Principal Component Analysis}\label{sec:pca}
\begin{figure}
	\includegraphics[width=\linewidth]{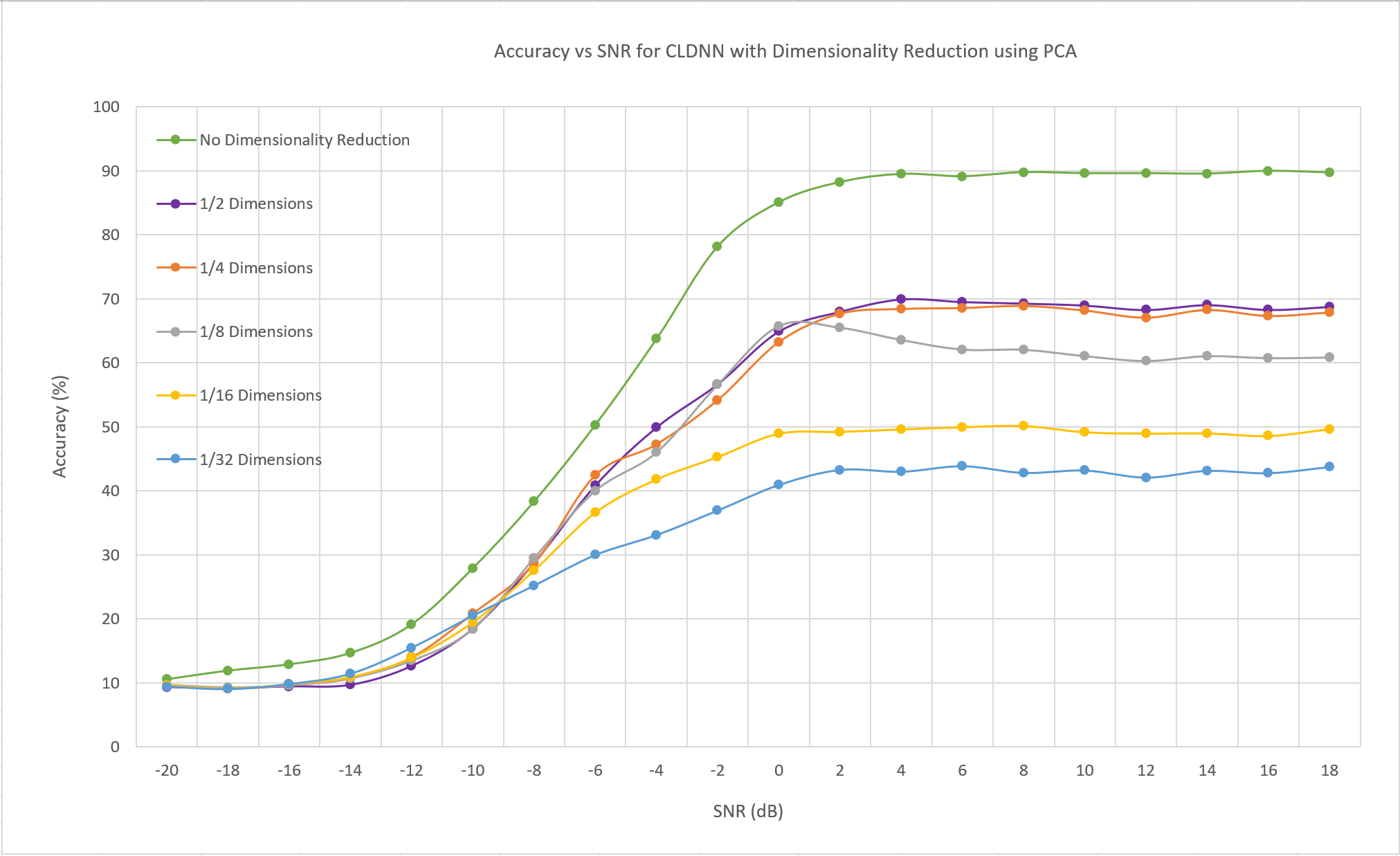}
	\caption{Classification accuracy vs. SNR using a CLDNN with different dimensionality reduction factors using PCA.}
	\label{fig:pca1}
\end{figure}
\begin{figure}
	\includegraphics[width=\linewidth]{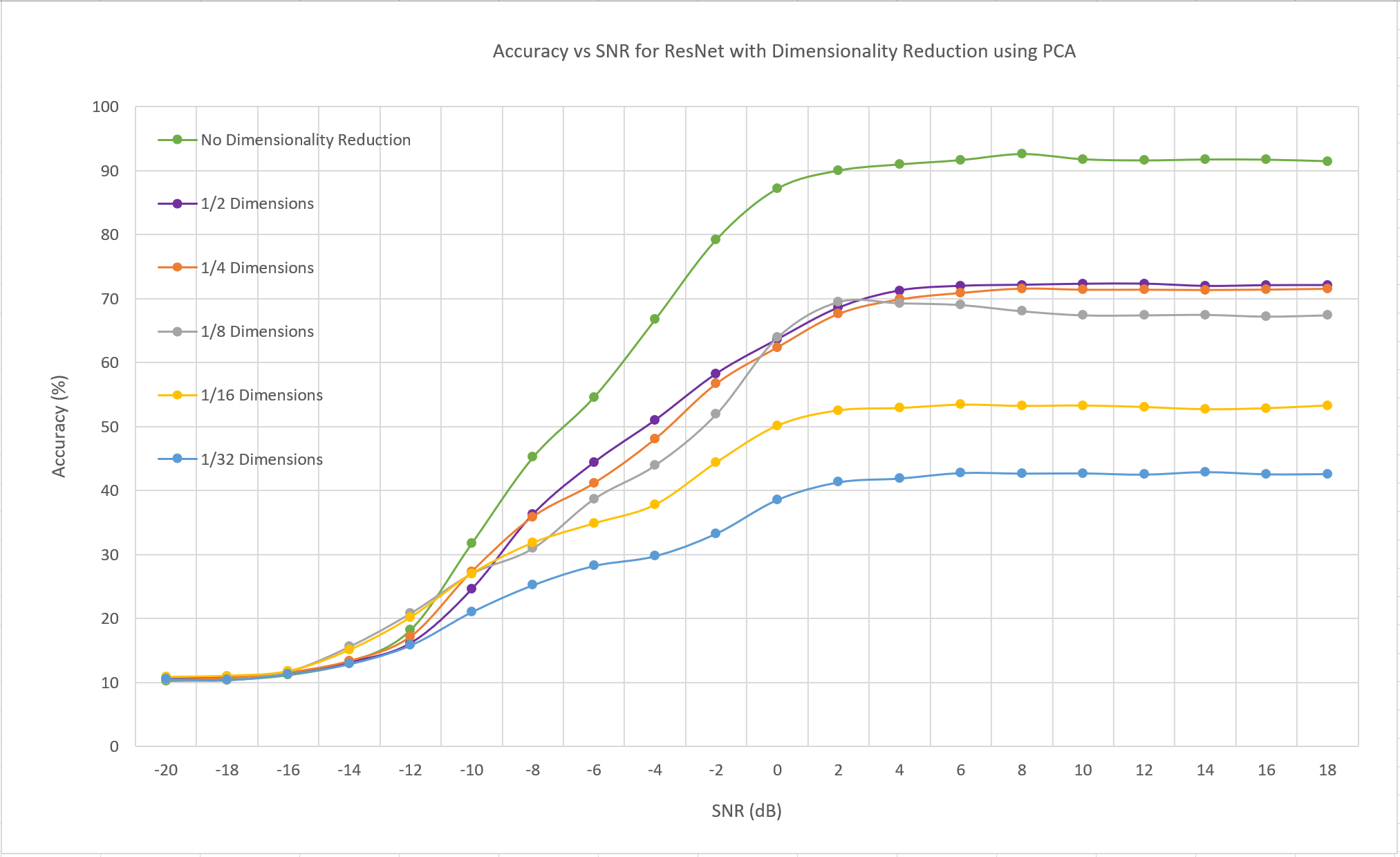}
	\caption{Classification accuracy vs. SNR using a ResNet with different dimensionality reduction factors using PCA.}
	\label{fig:pca2}
\end{figure}
\begin{figure}
	\includegraphics[width=\linewidth]{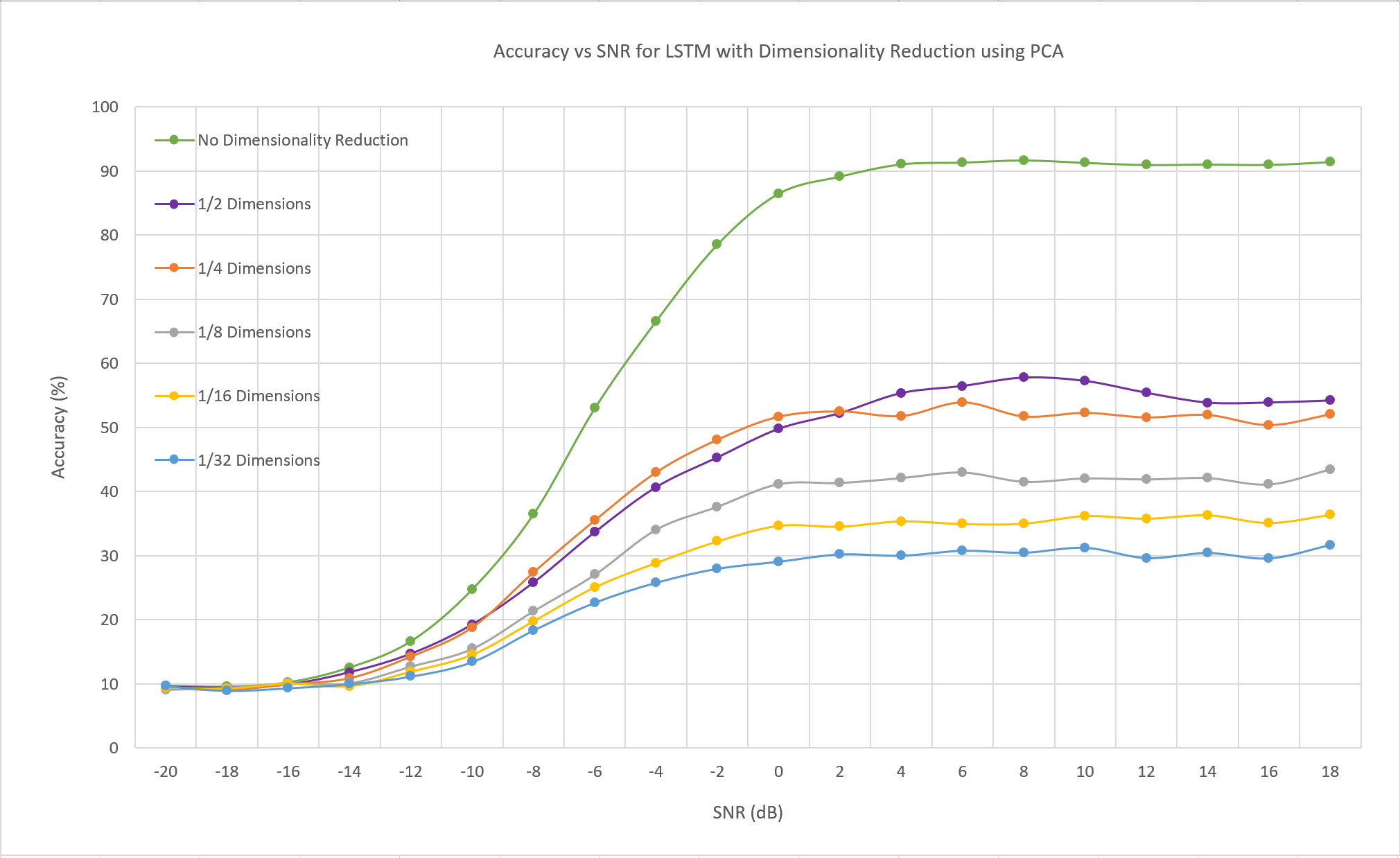}
	\caption{Classification accuracy vs. SNR using an LSTM with different dimensionality reduction factors using PCA.}
	\label{fig:pca3}
\end{figure}
Our first attempt is to use PCA~\cite{pca} to reduce the number of dimensions occupied by each of the input vectors. We perform PCA using all training input vectors, corresponding to the 10 modulation types. We find the basis of the reduced dimensional subspace based on the training data, and then project each of the test vectors on that same subspace. The results obtained by applying PCA to the input of the CLDNN, ResNet and LSTM architectures are given in Figs.~\ref{fig:pca1}, \ref{fig:pca2} and \ref{fig:pca3}, respectively. We first notice that for all considered architectures, the \textbf{training time drops in a linear fashion with the reduction of dimensions}. For example, reducing the dimensions by a factor of 2 leads approximately to halving the training time. We further make the following observations from the results:
\begin{enumerate}
    \item The performance of LSTM drops significantly when reducing the input dimensions. We believe that this is due to loss of temporal correlations that are strongly relevant to the classification task.
    \item The ResNet architecture is the most robust to dimensionality reduction using PCA, especially when reducing the dimensions by a factor of 8, which delivers an accuracy of approximately 70$\%$ at 2 dB.
    \item Interestingly, the accuracy curves are not necessarily monotonic with the SNR when reducing the input dimensions through PCA. This is most pronounced when reducing the dimensions by a factor of 8 for the input of the CLDNN architecture. The accuracy, in this case, drops significantly for SNR values higher than 0 dB. 
    \item It is not necessarily the case that the accuracy drops when reducing the input dimensions. We observe that we obtain almost identical results for all three architectures when reducing the input dimensions by factors of 2 and 4. We believe that this phenomenon occurs because even though dimensionality reduction may lead to losing relevant information needed for classification, it can also reduce overfitting by getting rid of task-irrelevant information.
\end{enumerate}

\subsection{Uniform Subsampling}\label{sec:subsampling}
\begin{figure}
	\includegraphics[width=\linewidth]{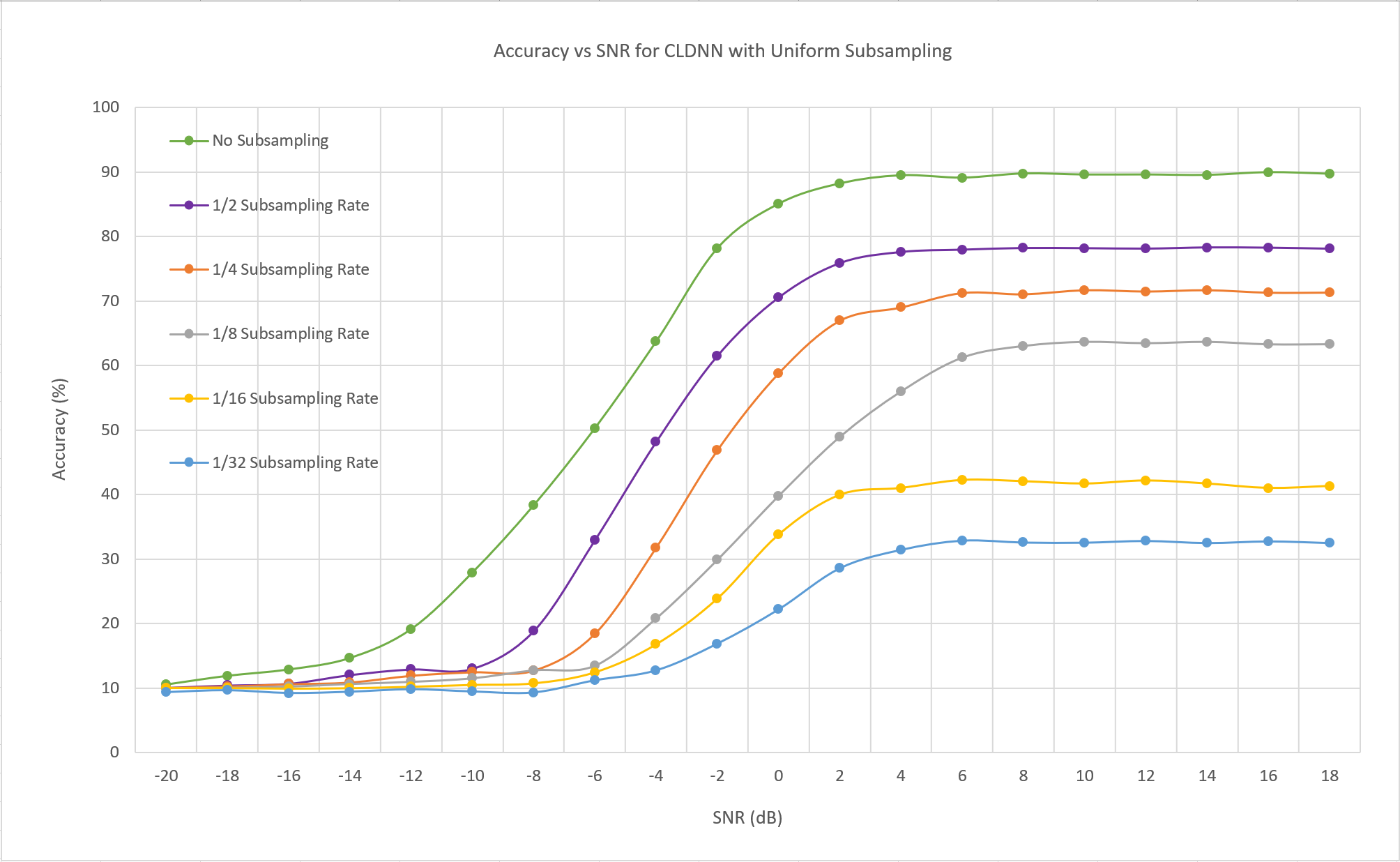}
	\caption{Classification accuracy vs. SNR using a CLDNN with different degrees of Uniform Subsampling.}
	\label{fig:uniform1}
\end{figure}
\begin{figure}
	\includegraphics[width=\linewidth]{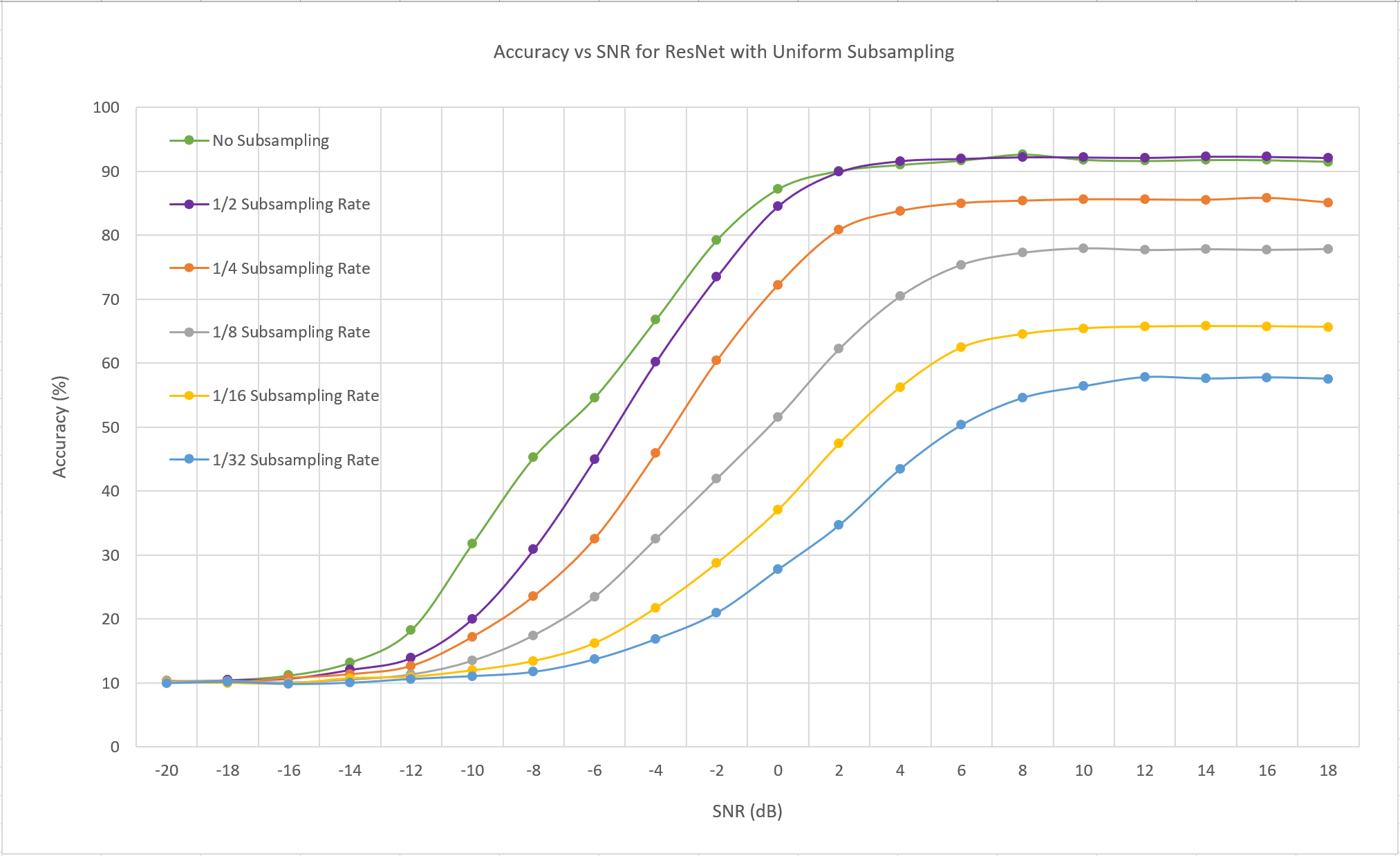}
	\caption{Classification accuracy vs. SNR using a ResNet with different degrees of Uniform Subsampling.}
	\label{fig:uniform2}
\end{figure}
\begin{figure}
	\includegraphics[width=\linewidth]{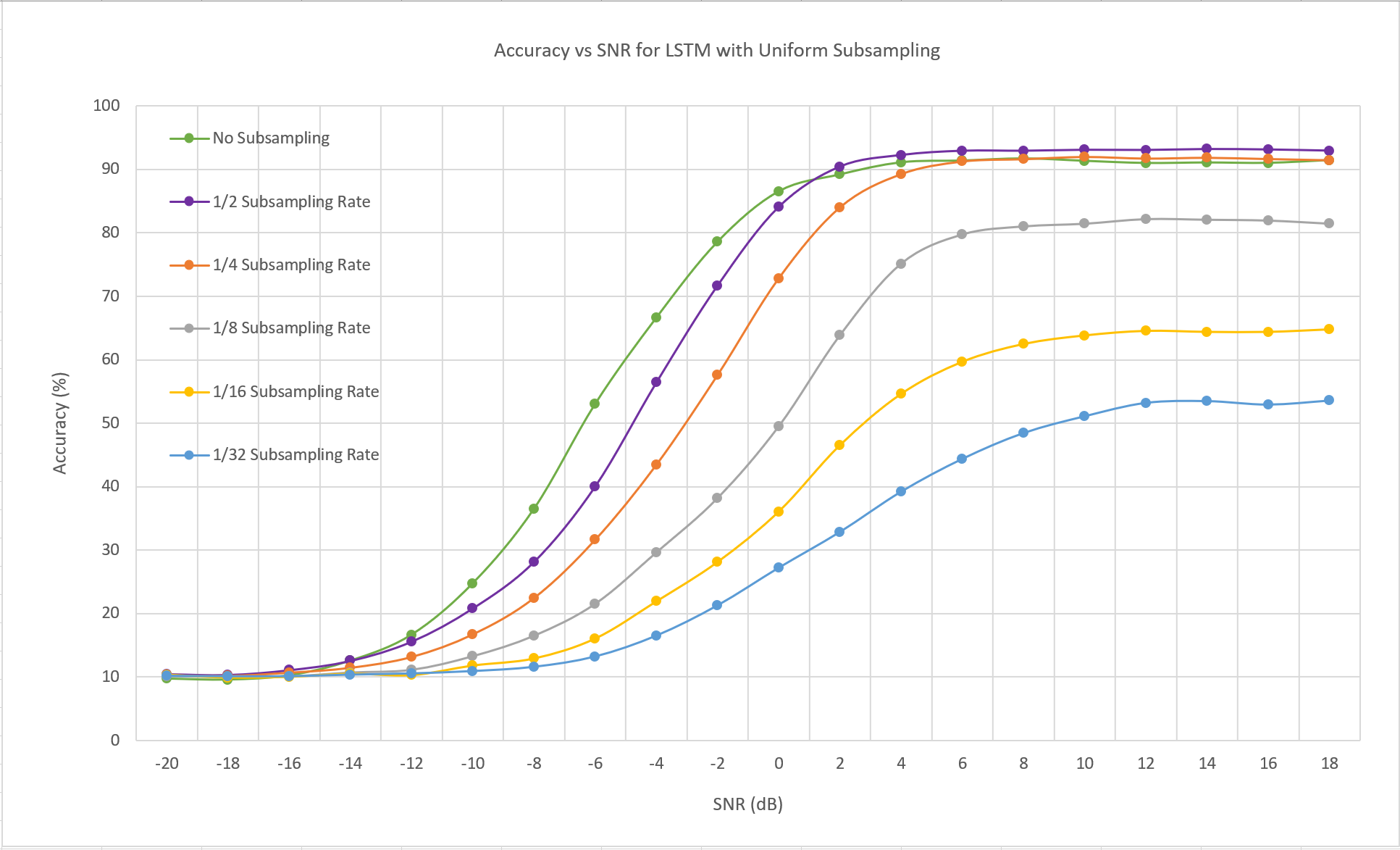}
	\caption{Classification accuracy vs. SNR using an LSTM with different degrees of Uniform Subsampling.}
	\label{fig:uniform3}
\end{figure}
Our second attempt is to use uniform subsampling~\cite{yonina-book} to accomplish the same task of dimensionality reduction as a means of reducing the training time of the candidate architectures. We sample the input vector at regular intervals and train the architectures based on the subsampled vector. The results obtained by uniform subsampling of the input of the CLDNN, ResNet and LSTM architectures are given in Figs.~\ref{fig:uniform1}, \ref{fig:uniform2} and \ref{fig:uniform3}, respectively. We notice that, like the results observed for the PCA experiments, the training time drops linearly with a drop in the number of dimensions of the input vector. For the best-considered architectures at high SNR - namely, LSTM and ResNet - we observe that uniform subsampling delivers superior performance to that of PCA. On the contrary, for the CLDNN and ResNet architectures that perform best at low SNR, PCA delivers better performance than uniform subsampling. As we will see in the rest of this section, this observation holds when comparing any of the subsampling methods considered in this work with PCA. This \textbf{attests to the excellent ability of PCA to combat high levels of random noise, while it also highlights its pitfalls at high SNR}, due to possibly losing structural information that is preserved with subsampling techniques. We further make the following observations from the uniform subsampling results:
\begin{enumerate}
    \item The performance of ResNet and LSTM increase when using half the samples at high SNR. For LSTM, the classification accuracy is higher with a quarter of the samples than that when using all the samples at high SNR. We believe that this is an effect of the oversampling of the training input (see Fig.~\ref{fig:fft}).
    \item The CLDNN architecture benefits from oversampling the input to obtain a high classification accuracy. This is evident in its performance when using half the samples, as a rapid drop in classification accuracy is observed. On the contrary, for ResNet and LSTM, the performance actually increases at high SNR as a result of the reduction in overfitting that is a side effect of training input oversampling.
    \item The performance when using half the samples by ResNet and LSTM is higher than that when using all the samples only at high SNR values (2 dB to 10 dB), whereas it is lower at low SNR values (-20 dB to 0 dB). This suggests that even the ResNet and LSTM architectures can \textbf{significantly benefit from oversampling at low SNR}. This advantage of oversampling at low SNR holds for all dimensionality reduction and subsampling techniques considered in this work.
\end{enumerate}

\subsection{Random Subsampling}\label{sec:subsampling}
\begin{figure}
	\includegraphics[width=\linewidth]{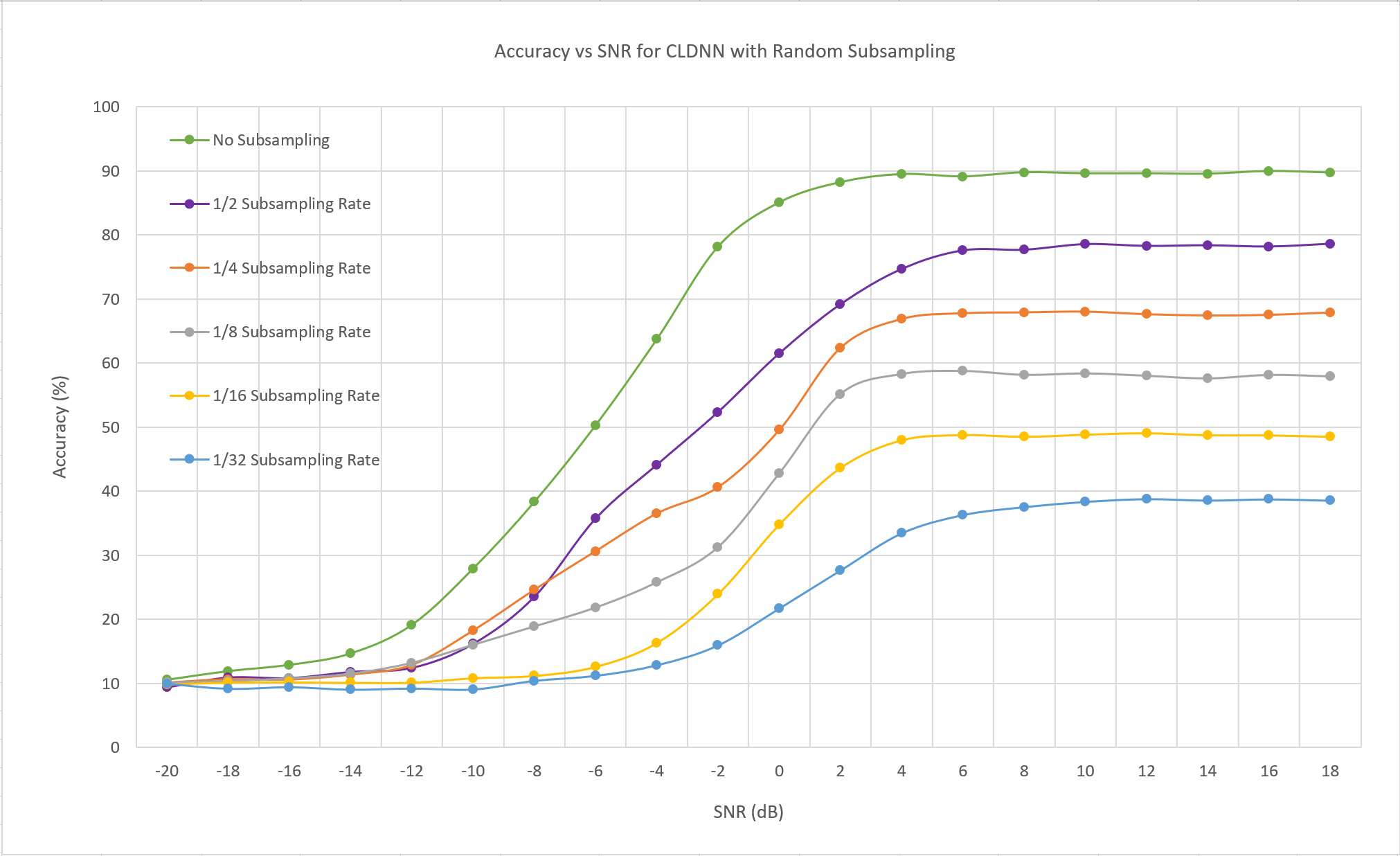}
	\caption{Classification accuracy vs. SNR using a CLDNN with different degrees of Random Subsampling.}
	\label{fig:random1}
\end{figure}
\begin{figure}
	\includegraphics[width=\linewidth]{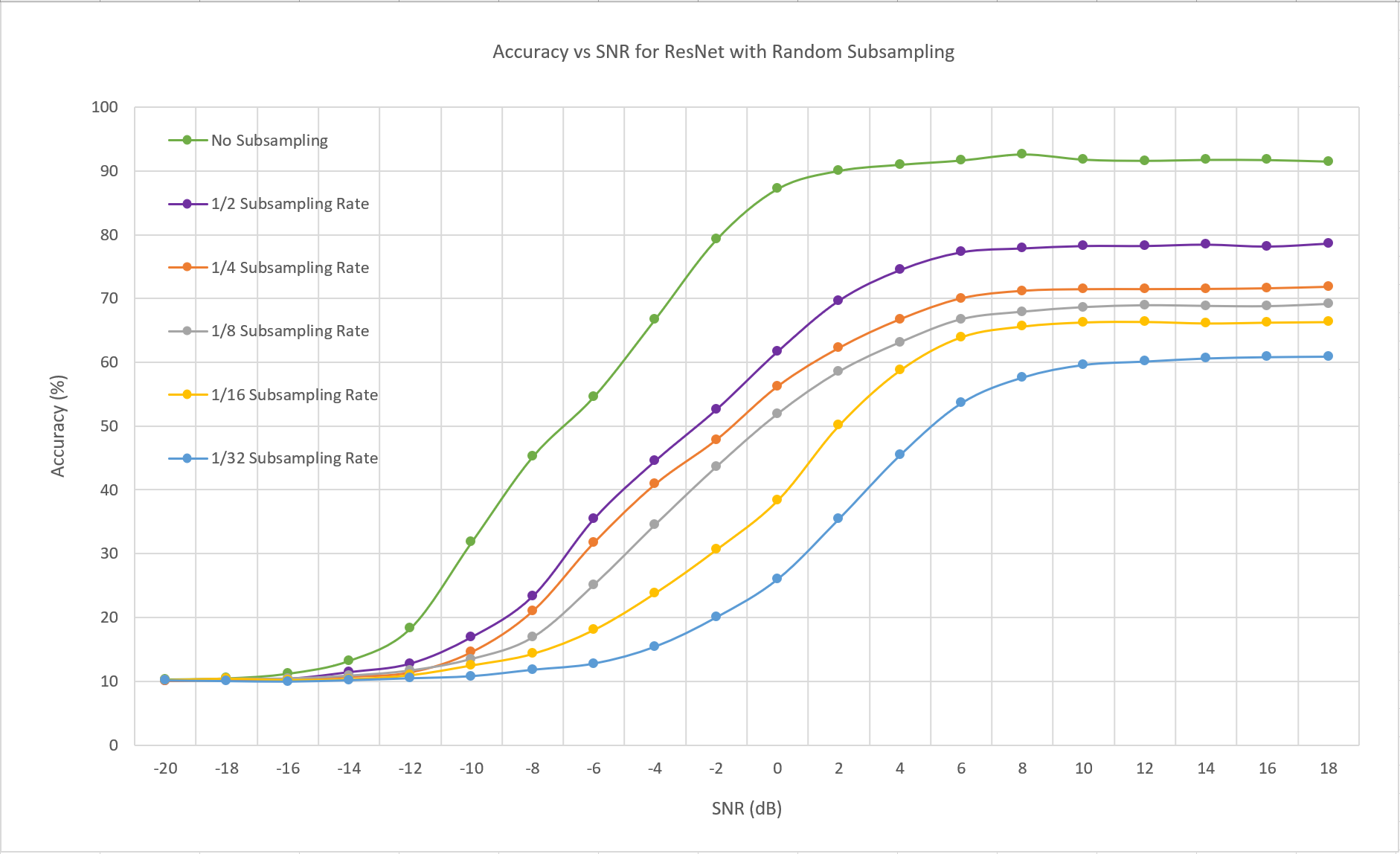}
	\caption{Classification accuracy vs. SNR using a ResNet with different degrees of Random Subsampling.}
	\label{fig:random2}
\end{figure}
\begin{figure}
	\includegraphics[width=\linewidth]{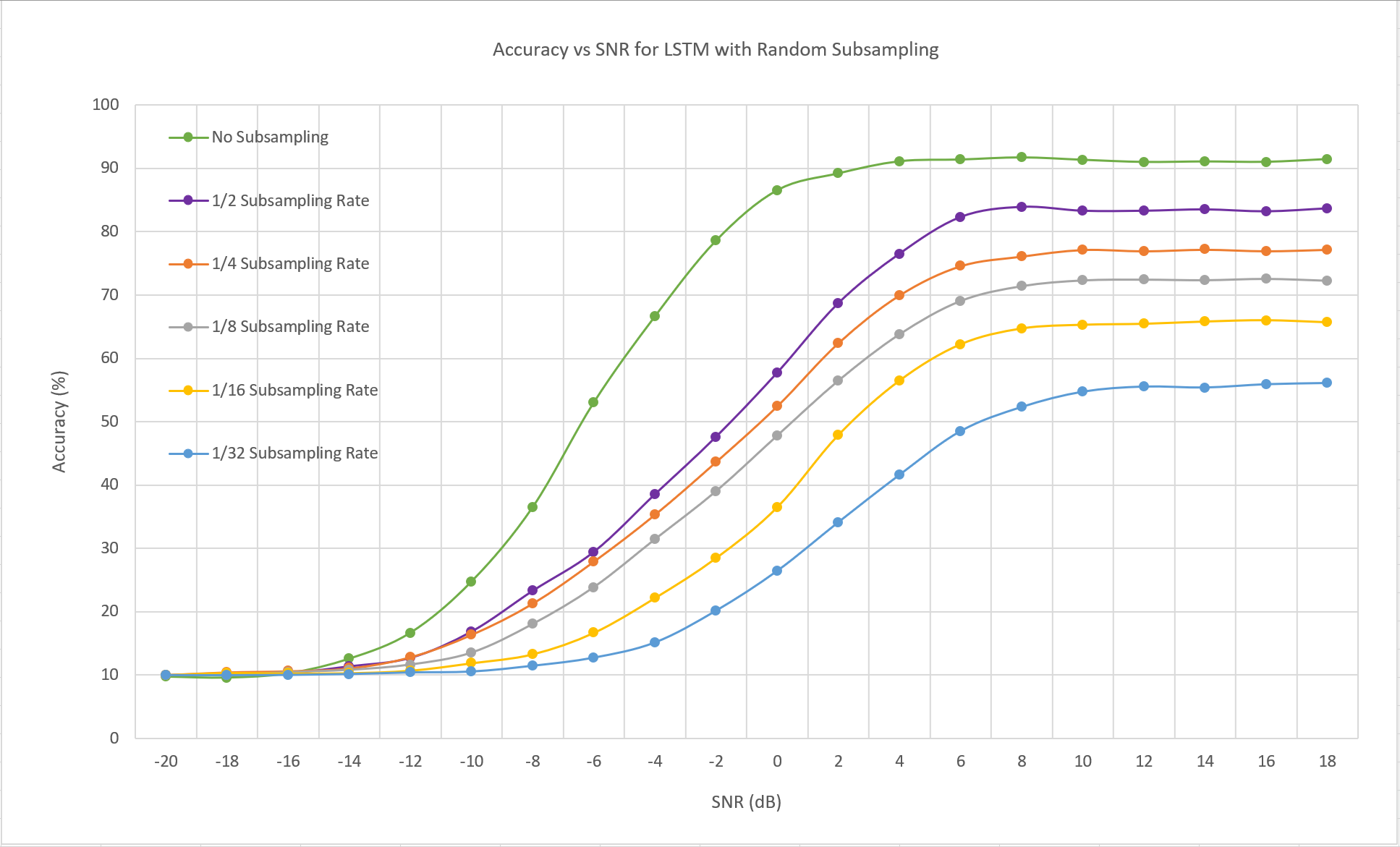}
	\caption{Classification accuracy vs. SNR using an LSTM with different degrees of Random Subsampling.}
	\label{fig:random3}
\end{figure}
Our third attempt to reduce the number of dimensions of the input vector is to use random subsampling~\cite{yonina-book}. Unlike uniform subsampling, where the input vector is sampled at uniform intervals, random subsampling attempts to sample the input vector at random intervals in time and train the architectures based on the subsampled vector. An important detail to note here is that the order in which the samples appear is maintained, which means that if two samples are collected at time $t$ and time $t+t_1$, where $t_1 > 0$, then the sample collected at $t+t_1$ must come after the sample collected at time $t$ in the resulting subsampled vector. Further, the randomization is performed only once to select a set of indices to subsample at, and then all training and testing vectors are subsampled at the same indices. The results obtained by random subsampling of the input of the CLDNN, ResNet and LSTM architectures are given in Figs.~\ref{fig:random1}, \ref{fig:random2} and \ref{fig:random3}, respectively. We note that the results are based on one random choice for the indices, as we obtained very similar results when trying multiple other choices. 


We observe that the uniformly sampled training data set leads to higher classification accuracy than the randomly sampled training data set when the resulting sampling rate is close to Nyquist ($1/8$ subsampling) or above ($1/4$ and $1/2$ subsampling). However, \textbf{random subsampling actually leads to higher classification accuracies for sampling rates well below the Nyquist rate} ($1/16$ and $1/32$ subsampling). This is consistent with the intuition in~\cite{yonina-book}, where typically sub-Nyquist strategies that are effectively non-uniform are superior.

\subsection{Magnitude Rank Subsampling}\label{sec:rank}
\begin{figure}
	\includegraphics[width=\linewidth]{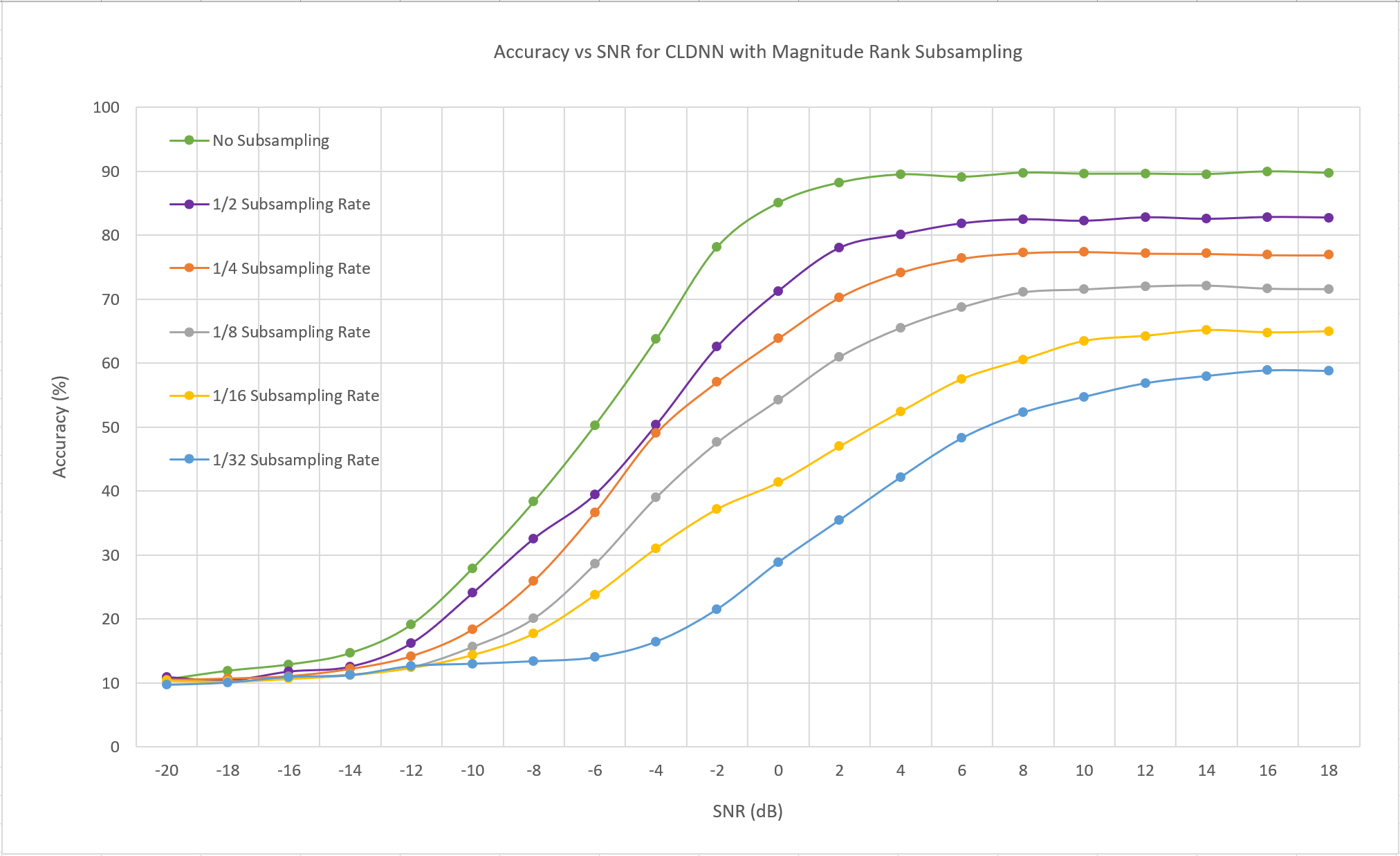}
	\caption{Classification accuracy vs. SNR using a CLDNN with different degrees of Magnitude Rank Subsampling.}
	\label{fig:rank1}
\end{figure}
\begin{figure}
	\includegraphics[width=\linewidth]{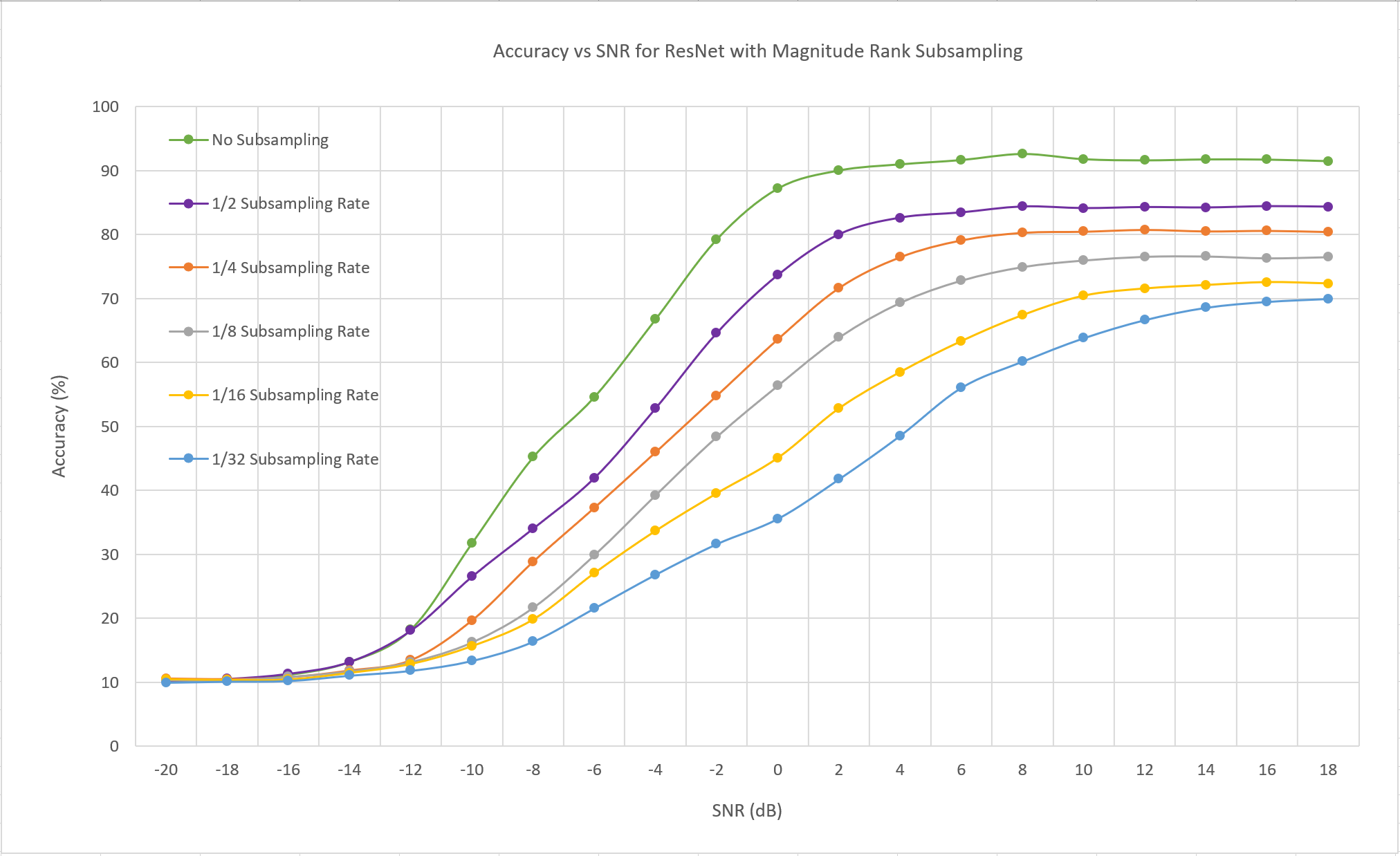}
	\caption{Classification accuracy vs. SNR using a ResNet with different degrees of Magnitude Rank Subsampling.}
	\label{fig:rank2}
\end{figure}
\begin{figure}
	\includegraphics[width=\linewidth]{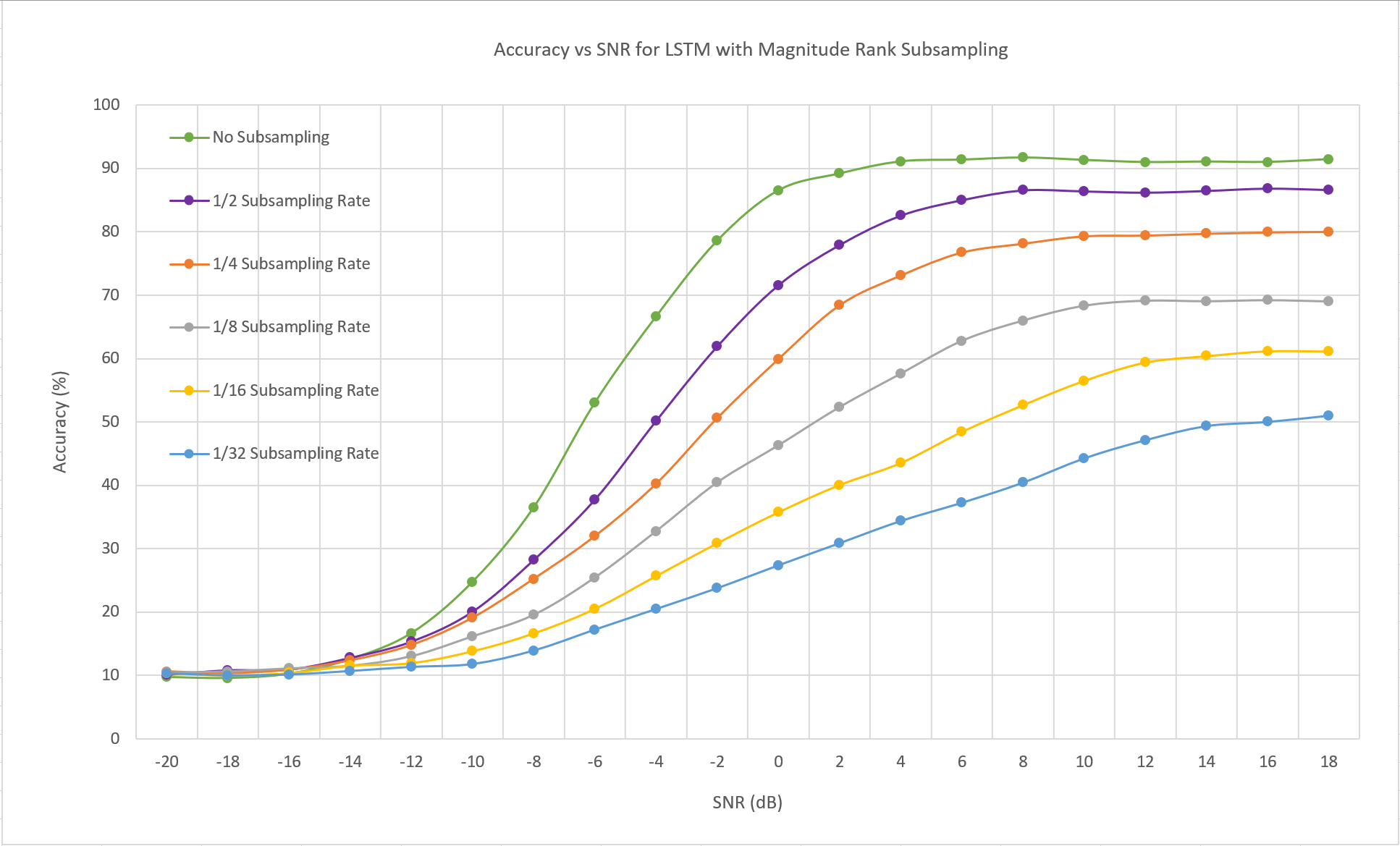}
	\caption{Classification accuracy vs. SNR using an LSTM with different degrees of Magnitude Rank Subsampling.}
	\label{fig:rank3}
\end{figure}
Inspired by the sub-Nyquist rate sampling techniques discussed in ~\cite{yonina-book}, we present our fourth and final attempt to perform dimensionality reduction of the input vectors as a means of reducing the training time. We use a magnitude-based subsampling, where first the real and imaginary parts of the samples are used to calculate the magnitudes of the samples. The samples corresponding to each vector are then ranked in the descending order of their magnitudes with rank 1 belonging to the sample with the largest magnitude. The top samples with the highest magnitudes are collected based on the subsampling rate and are rearranged back in the sequence that they were present in as observed in the original data set, which is similar to the maintenance of the order in which the samples appear as seen in the case of random subsampling. The results obtained by the described magnitude rank subsampling of the input of the CLDNN, ResNet and LSTM architectures are given in Figs.~\ref{fig:rank1}, \ref{fig:rank2} and \ref{fig:rank3}, respectively. We observe the following:
\begin{enumerate}
    \item At a sampling rate close to the Nyquist rate ($1/8$ subsampling) and above ($1/4$ and $1/2$ subsampling), magnitude rank subsampling performs worse than uniform subsampling and better than random subsampling in terms of classification accuracies for all three considered architectures. 
    \item \textbf{Magnitude rank subsampling performs significantly better than both uniform and random subsampling when operating well below Nyquist rates ($1/16$ and $1/32$ subsampling), with the exception of the LSTM architecture}. Again, this is consistent with the intuition discussed in~\cite{yonina-book} for sub-Nyquist sampling.
    \item The relatively poor performance of magnitude rank subsampling with the LSTM architecture compared to the other two architectures can be attributed to the loss of temporal correlations that are strongly relevant to the classification task, which could have relied on samples with lower magnitudes.
    \item The degradation in accuracy due to lowering the sampling rate seems to be closer to \emph{linear} than PCA and uniform and random subsampling. In particular, this degradation is considerably mild at high SNR.
\end{enumerate}

\subsection{Discussion: Training Time Reduction}\label{sec:time}
The objective of the dimensionality reduction and subsampling techniques is to achieve a reduction in the training times of the architectures, in order to facilitate online training. The following are the trends observed in the training times of the three networks:
\begin{enumerate}
    \item In all considered cases of dimensionality reduction and subsampling, a \textbf{linear drop in training time is observed with a drop in the number of dimensions of the input vector}. More precisely, the ratio of the training time per epoch before the application of PCA or subsampling, to the training time per epoch of after PCA or subsampling, is approximately the same as the ratio of dimensions of the input vector before dimensionality reduction, to the dimensions of the input vector after dimensionality reduction.
    \item The above-mentioned training time ratio is slightly higher for the CLDNN and the ResNet architectures when the ratio of the number of dimensions before and after PCA or subsampling is close to or higher than the Nyquist rate, i.e., at $1/8$, $1/4$, and $1/2$ reduction in dimensions. For the CLDNN, the training time ratios are approximately $0.131$, $0.28$, and $0.57$, respectively, whereas for the ResNet, the training time ratios are $0.129$, $0.27$, and $0.51$, respectively.
    \item The second observation above is not applicable to the LSTM architecture, even for the $1/2$ dimensionality reduction and subsampling rate. We believe that this is because the input layer of the LSTM consists of a higher proportion of the overall number of neurons in the architecture, compared to input layers of the CLDNN and ResNet architectures. 
\end{enumerate}

\section{SNR Selection}\label{sec:snr}
In this section, we consider training the CLDNN, ResNet and LSTM architectures identified in Section~\ref{sec:architectures} using only the data sets corresponding to one or two representative SNR values, instead of all available 20 SNR values. As a result, the training time is significantly reduced. We further show how certain choices for these \emph{training SNR values} result in negligible losses in the classification accuracy.
\subsection{Single SNR Selection}\label{sec:singlesnr}
\begin{figure}[htb]
	\includegraphics[width=\linewidth]{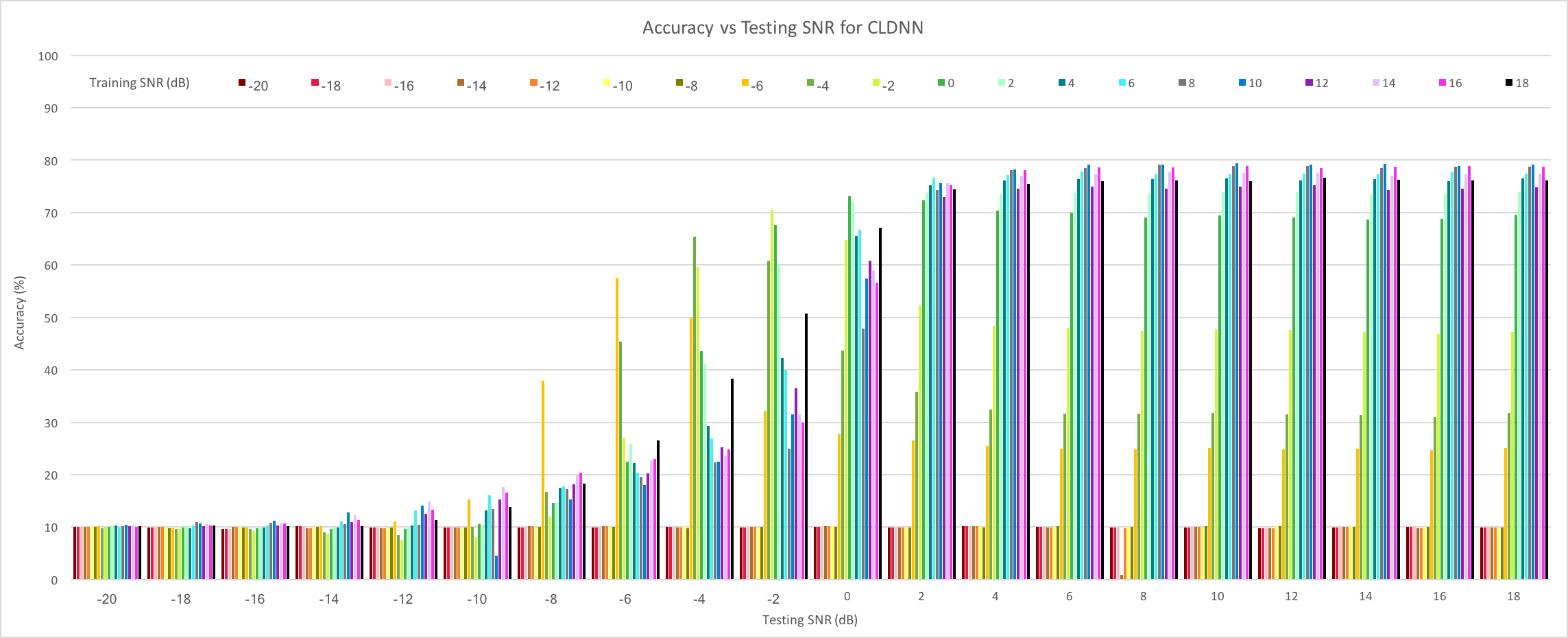}
	\caption{Single SNR Selection Result of CLDNN.}
	\label{fig:single1}
\end{figure}

We first consider training each architecture with a data set collected at a single SNR value. For the CLDNN architecture, the result of classification accuracy versus different training SNR values is shown in Fig. \ref{fig:single1}. Training only with 10 dB data gave the best performance. The training time was reduced to 3 seconds per epoch compared to 58 seconds per epoch before. We also note that training with a single SNR in the intermediate range between -8 dB and 0 dB produced high classification accuracy for the testing data sets at corresponding intermediate SNR values. 

\begin{figure}[htb]
	\includegraphics[width=\linewidth]{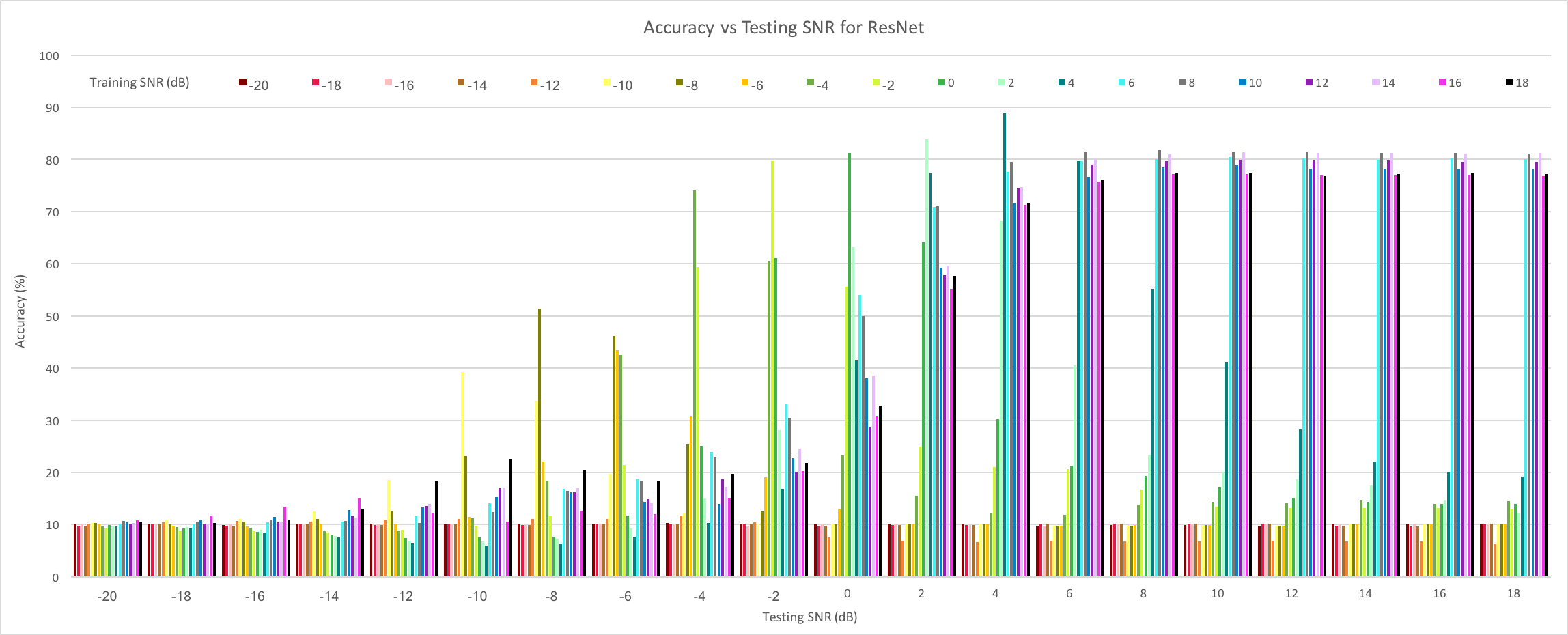}
	\caption{Single SNR Selection Result of ResNet.}
	\label{fig:single2}
\end{figure}

Fig. \ref{fig:single2} shows the ResNet performance for training with individual SNR data. Training with high SNR data produced better overall accuracy and the 8 dB training data set led to the highest overall classification accuracy. Training time was reduced to only 2 seconds per epoch using all 3 GPUs compared to 38 seconds per epoch using all 3 GPUs. 

\begin{figure}[htb]
	\includegraphics[width=\linewidth]{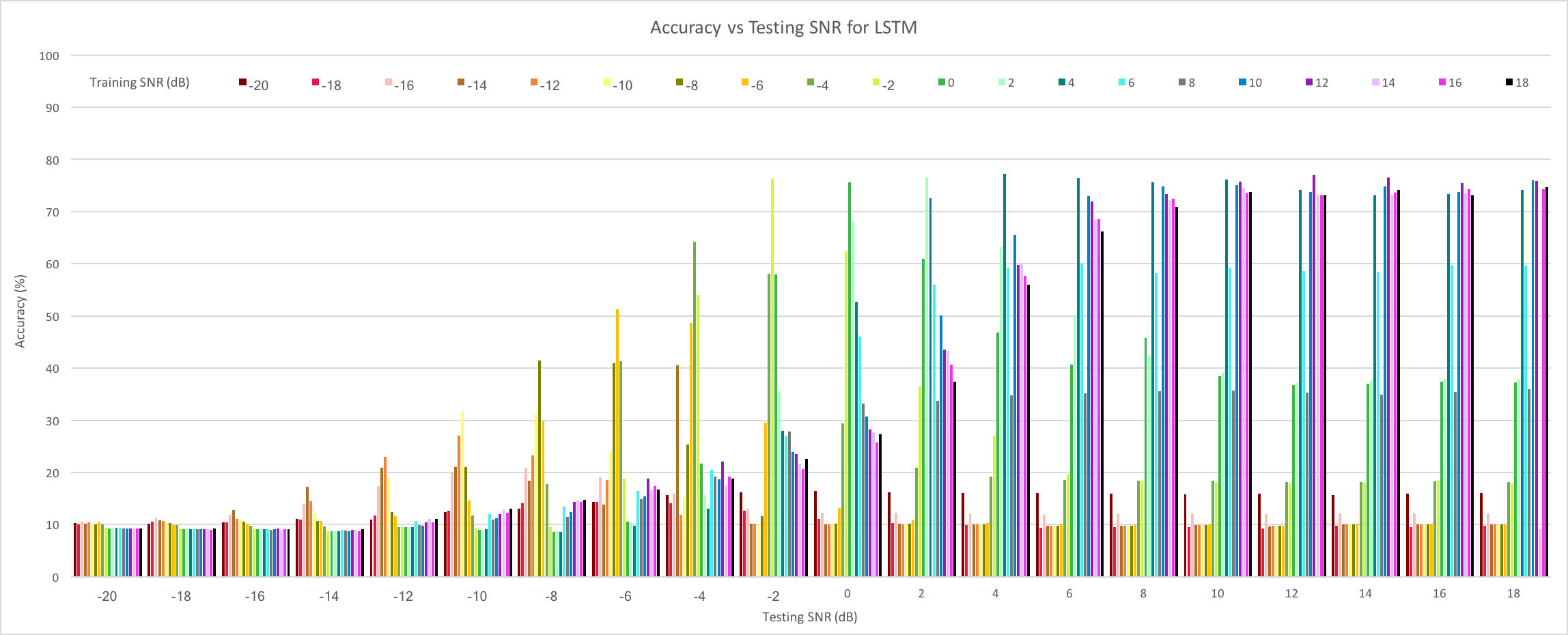}
	\caption{Single SNR Selection Result of LSTM.}
	\label{fig:single3}
\end{figure}

Fig. \ref{fig:single3} shows the model performance of LSTM for individual SNR training. Training with 4 dB only led to the highest overall accuracy. Training time reduced to 12 seconds per epoch compared to 222 seconds per epoch using all 3 GPUs. 

Based on results for all three considered networks, we note that training with high SNR data produces the highest average testing accuracy over all considered SNR values; in particular, it results in significantly higher accuracy for high SNR testing data. Training with very low SNR data (below -10 dB) does not seem beneficial at all. However, training at low SNR values in the range between -10 dB and 0 dB produces the highest accuracy values on testing data in the same SNR range, but not for higher SNR testing data. 

\subsection{Uniform Random Selection}\label{sec:uniformsnr}
\begin{figure}[htb]
	\includegraphics[width=\linewidth]{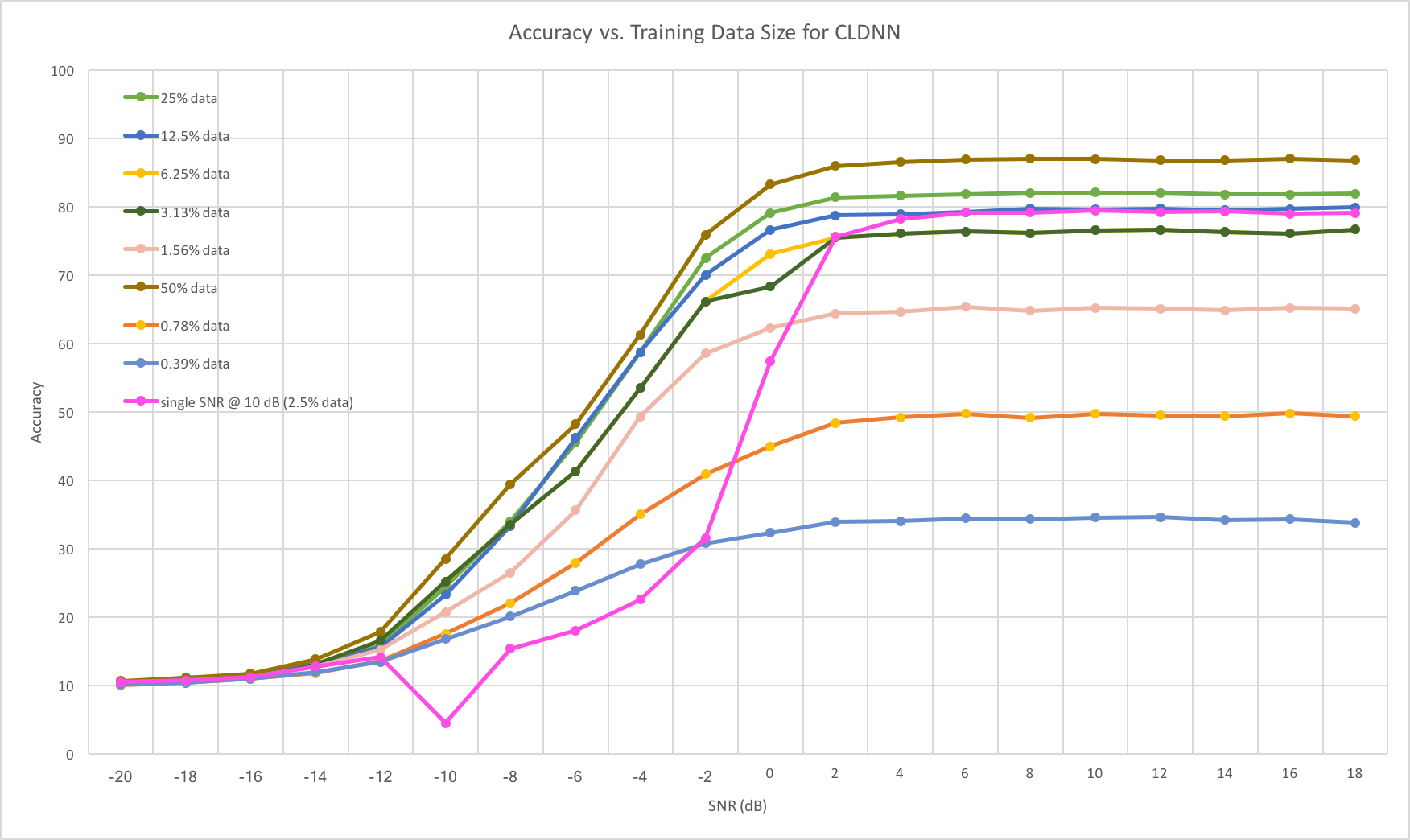}
	\caption{Classification accuracy vs. training data size using a CLDNN.}
	\label{fig:size1}
\end{figure}

\begin{figure}[htb]
	\includegraphics[width=\linewidth]{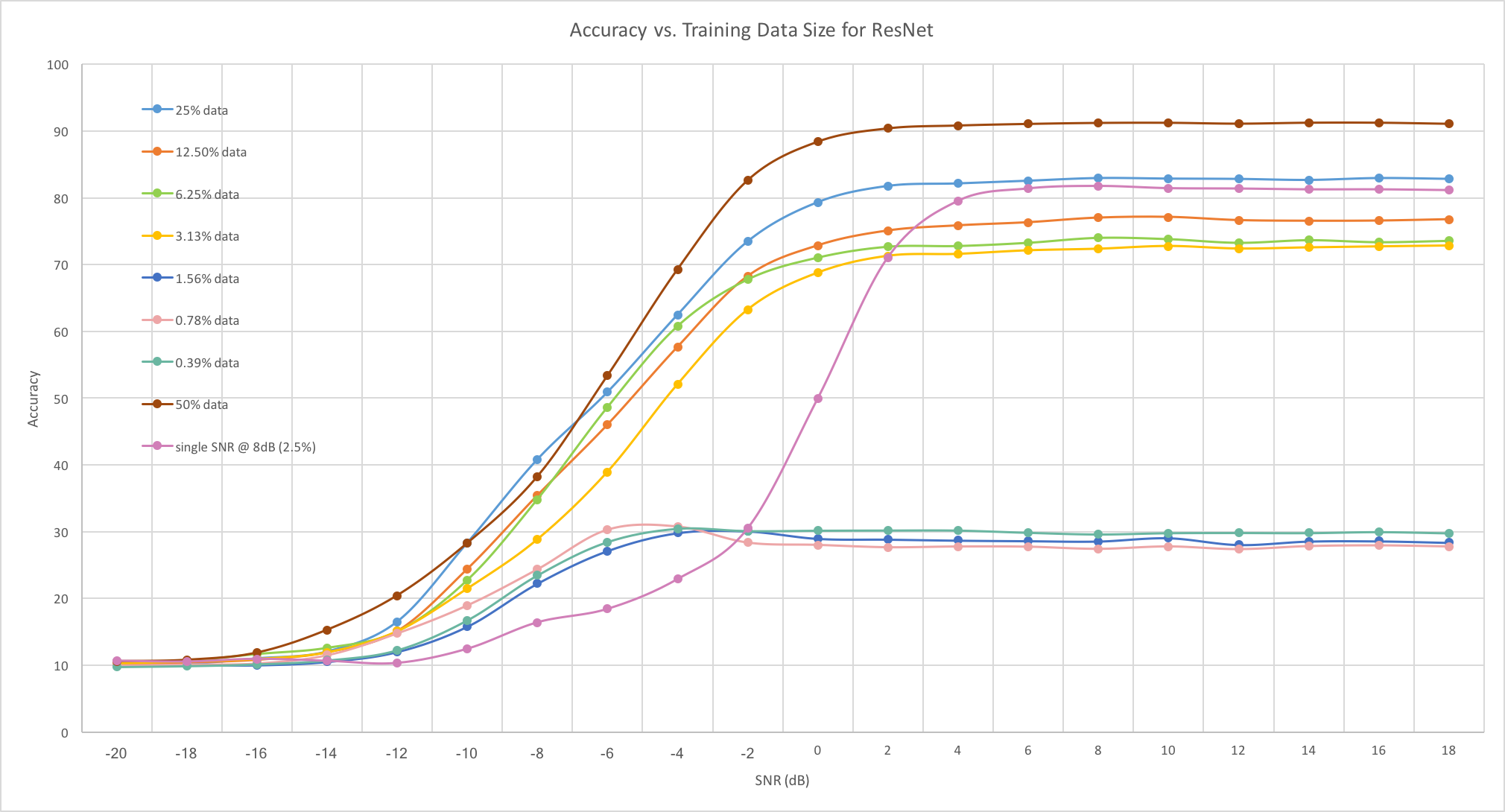}
	\caption{Classification accuracy vs. training data size using a ResNet.}
	\label{fig:size2}
\end{figure}

\begin{figure}[htb]
	\includegraphics[width=\linewidth]{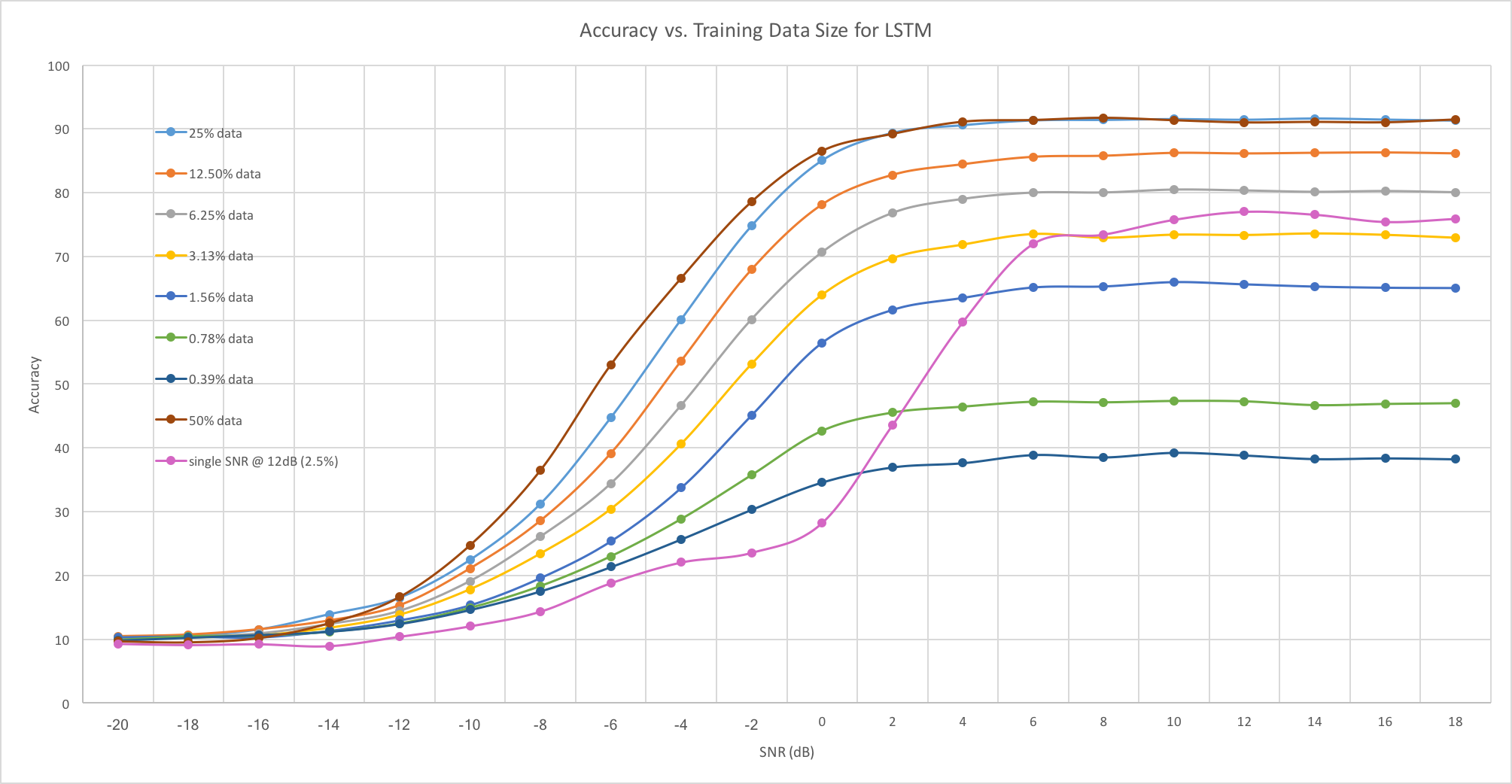}
	\caption{Classification accuracy vs. training data size using an LSTM.}
	\label{fig:size3}
\end{figure}

In order to evaluate the effectiveness of choosing representative SNR training values, we next experiment with randomly selecting a training data set, that is equally split among all the 20 SNR values. We select the training data set by combining equi-sized sets, that are randomly selected from each of the 20 SNR values. Figs. \ref{fig:size1}, \ref{fig:size2}, and \ref{fig:size3} show the classification accuracies for various training data sizes for CLDNN, ResNet, and LSTM, respectively. Note that in the original setup, we used 50\% of the data set for training. The percentages of data showed on the plots correspond to percentages of the entire data set.

We also compare the obtained results for each of the three architectures with those obtained by training using the training data set at the single SNR value that gives the highest average classification accuracy (identified in Section~\ref{sec:singlesnr}). Since we have a total of 20 SNR data sets, training using a single SNR corresponds to using only $\frac{50\%}{20}=2.5\%$ of the whole data set for training. We note that using the representative SNR for training always gives higher classification accuracy at high testing SNR values for all three architecture, than using a uniform data set with the bigger size of $\frac{50\%}{16}=3.125\%$ of the whole data set. A similar phenomenon occurs when testing at low SNR values, and choosing a representative low SNR value for training, which indicates that \textbf{if we have a good estimate of the range of SNR values, at which the classifier would operate, then speeding up the training time using only the training set at representative SNR values is better than uniformly sampling the training set across all SNR values}. However, if such an estimate is absent, then the uniform selection is preferred, as it delivers better performance when there is a significant mismatch between the trained and testing SNR values, as can be seen in Figs. \ref{fig:size1}, \ref{fig:size2}, and \ref{fig:size3} at low SNR values.

We found all three networks to retain relatively high accuracy (above 70\% at high SNR) as long as the training data size is at least 3.125\% of the entire data set size. Further, the \textbf{CLDNN and LSTM architectures - that capture long-term dependencies - are resilient to more aggressive reductions in the training data set size}. Finally, it is worth noting that we found the training time to drop in a linear fashion with the reduction in the size of the training data set.  

\subsection{Pairs of SNR Selection}\label{sec:pairsnr}
We notice from the results in Section~\ref{sec:singlesnr} that there are roughly two SNR ranges, according to the optimal training SNR value; the range of high SNR values, and the range of mildly low SNR values. Based on this observation, we use the data sets corresponding to a pair of SNR values in this section, and investigate the existence of a pair that gives a high classification accuracy over a wide range of SNR values.

\begin{figure}[htb]
	\includegraphics[width=\linewidth]{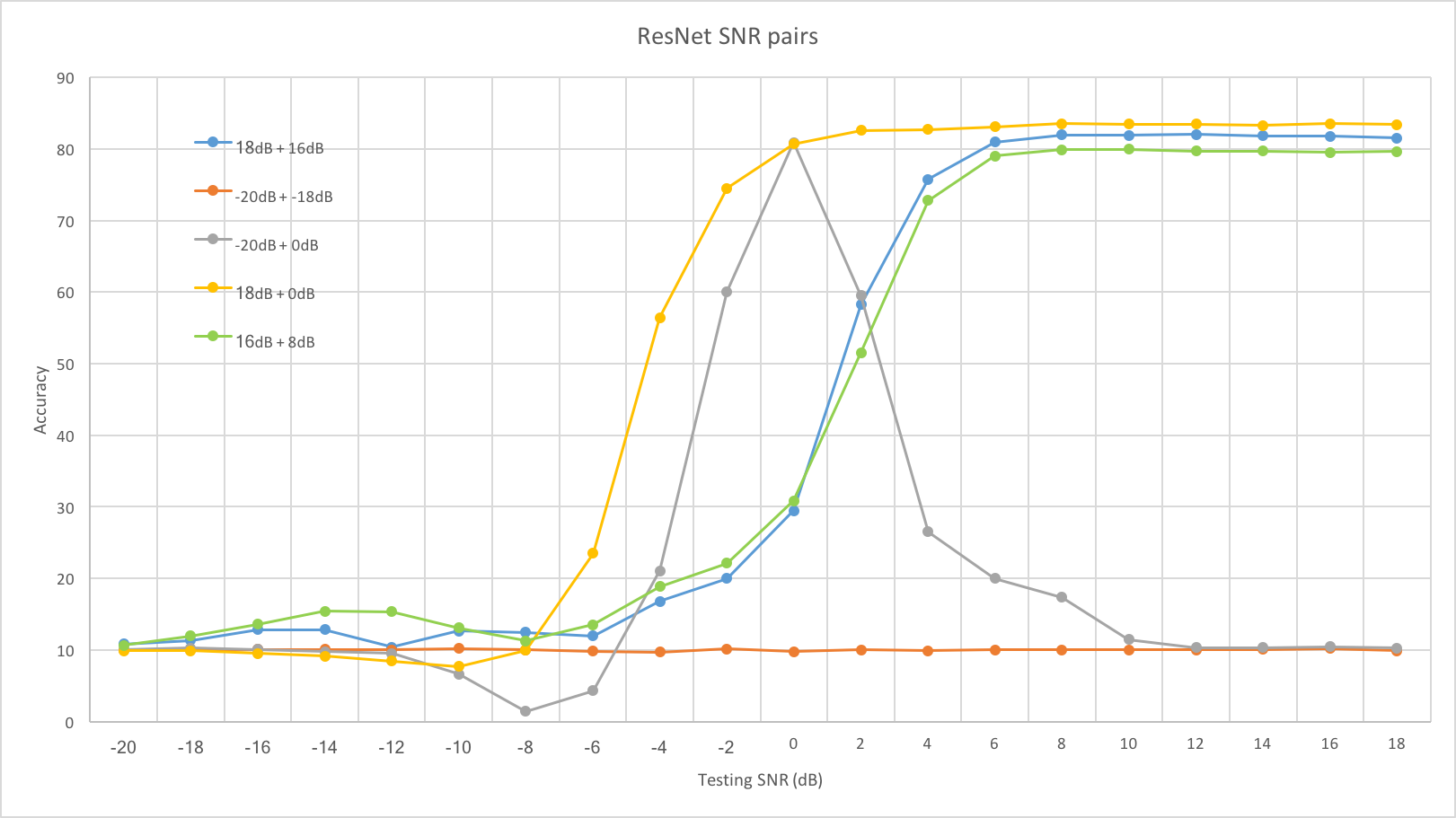}
	\caption{Classification accuracy vs. SNR using a ResNet with SNR pairs selection.}
	\label{fig:pair1}
\end{figure}

Fig. \ref{fig:pair1} shows the SNR pair selection results for ResNet. We can see from the plot that using 18 dB and 0 dB for training produced the best overall accuracy. Training with the two highest SNR data, 18 dB and 16 dB, gave high accuracy only for high SNR testing data. The combination of 16 dB and 8 dB was also tested because of their high individual SNR selection performance. Training with only low SNR data yielded an accuracy of only around 10\%. This means that the training data contains too much noise and our model was not able to identify the patterns for each modulation scheme. Training with -20 dB and 0 dB resulted in high accuracy only around 0 dB, which suggested that the model only relies on the 0 dB training data. \textbf{It is interesting to note that one pair of SNR values leads to achieving the highest classification accuracy - among all tested pair choices - for a wide range of SNR values from -6 dB to 18 dB}.

\begin{figure}[htb]
	\includegraphics[width=\linewidth]{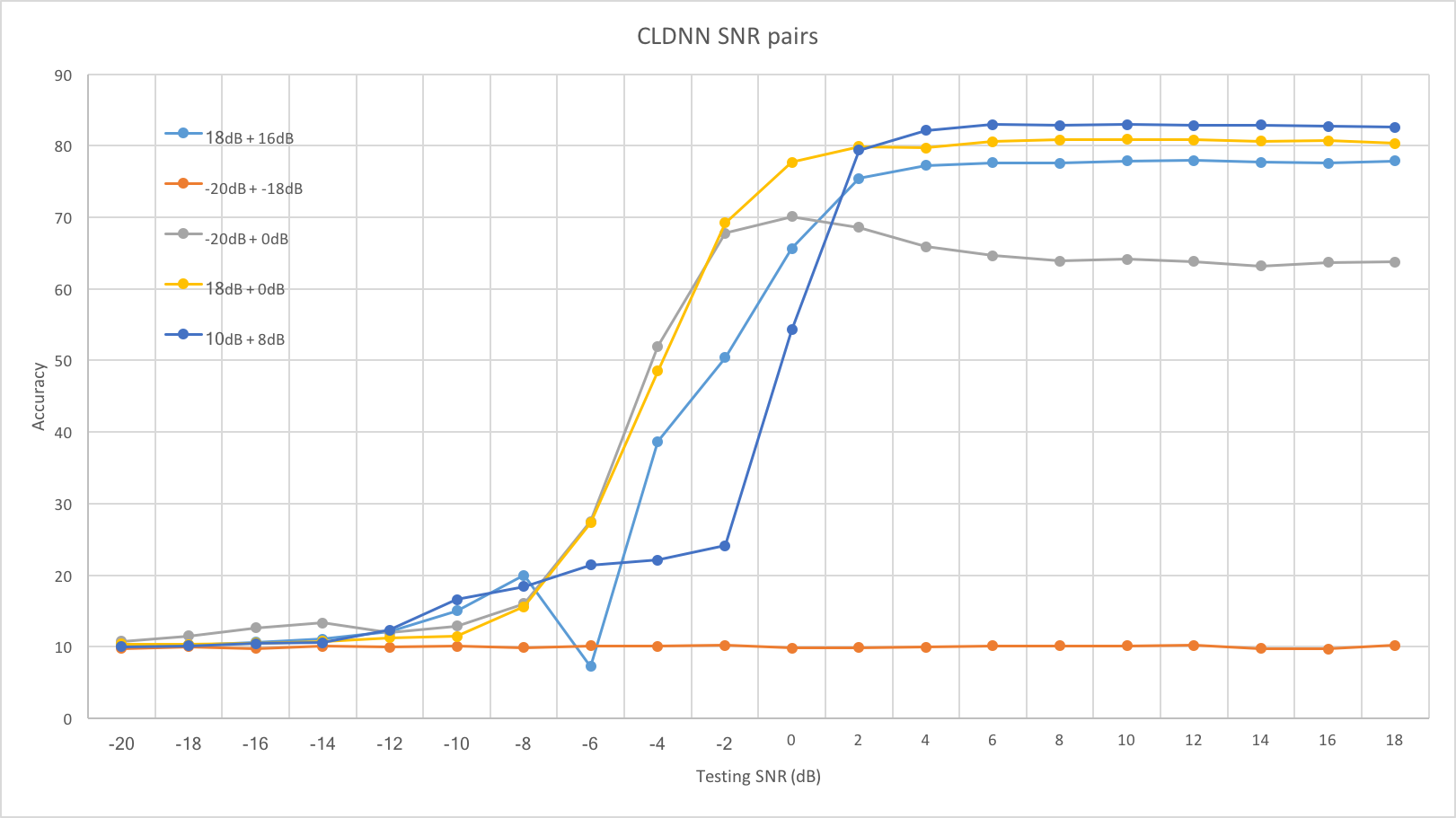}
	\caption{Classification accuracy vs. SNR using a CLDNN with SNR pairs selection.}
	\label{fig:pair2}
\end{figure}

Fig. \ref{fig:pair2} shows the result for SNR pair selection for CLDNN. Training with 10 dB and 8 dB gave us the highest accuracy for high SNR testing. These two specific SNR values were selected because of their high individual SNR selection performance. Training only with the two lowest SNR data yielded low accuracy, similar to the results for the ResNet. Training with 18 dB and 0 dB resulted in better accuracy for low SNR values between -6 dB and  0 dB, compared to training with a pair of high SNR data. The model trained using -20 dB and 0 dB was able to retain an accuracy above 60\% even for high SNR testing, different from the results for the ResNet, \textbf{which poses the question of whether long-term dependencies - captured by the LSTM layer - could enable distilling useful information from low SNR data sets}. However, the accuracy curve still contains a decaying shape. 

\begin{figure}[htb]
	\includegraphics[width=\linewidth]{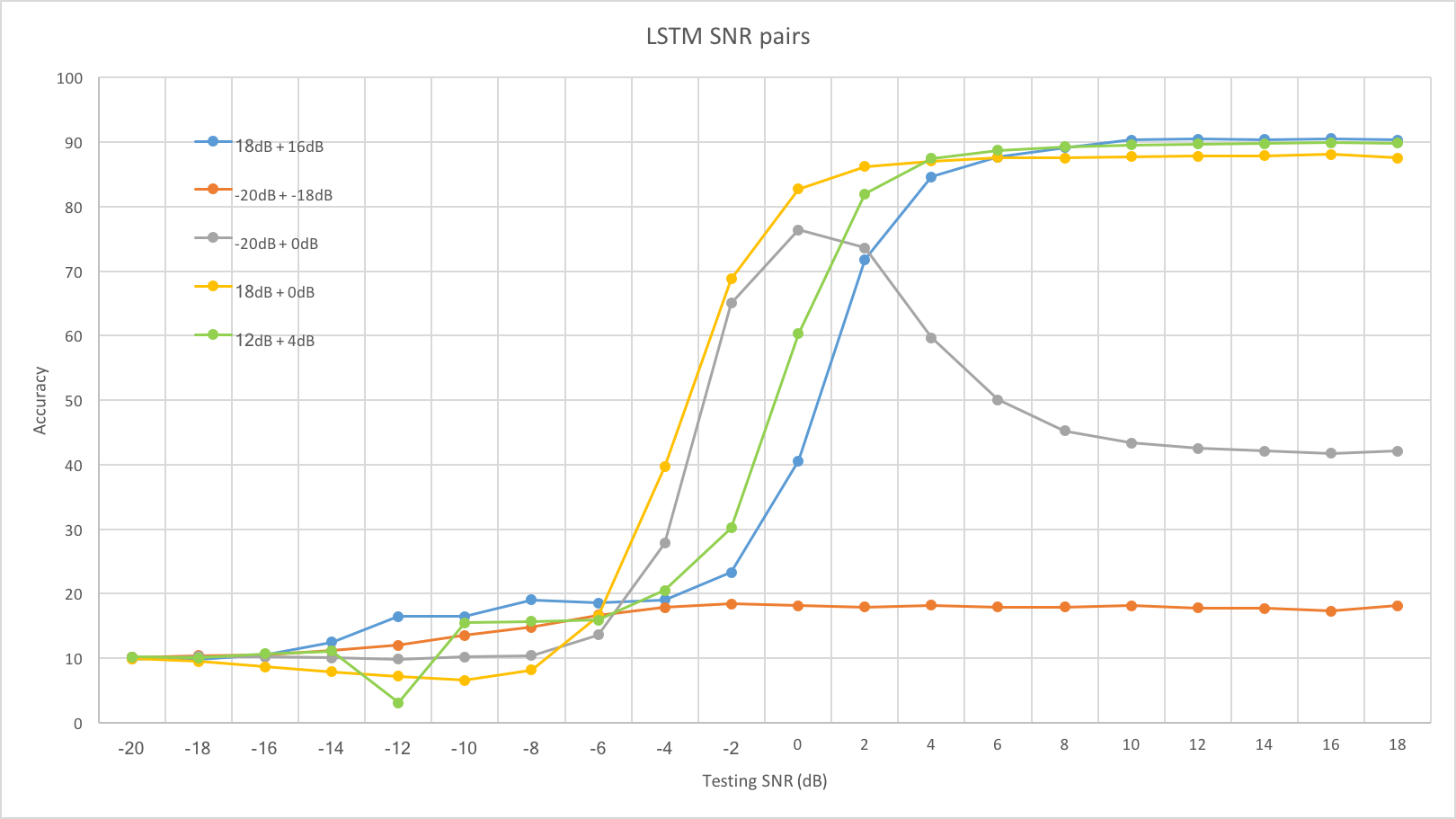}
	\caption{Classification accuracy vs. SNR using an LSTM with SNR pairs selection.}
	\label{fig:pair3}
\end{figure}

Fig. \ref{fig:pair3} shows the results of the SNR pair selection for the LSTM architecture. High testing accuracy around high SNR occurred for two pairs of high SNR training data, 18 dB $+$ 16 dB and 12 dB $+$ 4 dB. 12 dB and 4 dB were selected for their strong individual SNR selection performance. \textbf{Interestingly, the loss in accuracy for high SNR testing is less than 2\% for LSTM, while the training time reduced to 23 seconds per epoch compared to 222 seconds per epoch in the original setup}. Training with 18 dB and 0 dB still gave us higher accuracy in the 0 dB to -6 dB range compared to other pairs of training data. Training with -20 dB and 0 dB produced high accuracy at 0 dB, but the accuracy decayed for higher SNR's. Overall, training with 18 dB and 0 dB yielded the highest average accuracy across all tested SNR values, which is consistent with the insight that training with a representative SNR value from each range yields the best results. Training with a pair of high SNR data produces high accuracy for high SNR testing only. Training with -20 dB and 0 dB created an accuracy spike around 0 dB, but then a decaying curve for higher SNR values. By training with 18 dB and 0 dB, we could get similar performance compared to training with 50\% of the data across the range of tested SNR values, while reducing the training time by approximately 90\%. 

\section{Discussion}\label{sec:discussion}
\subsection{Which Algorithm to Choose}
In Section~\ref{sec:architectures}, we presented five different deep neural networks; all of them deliver higher classification accuracy than the CNN of~\cite{conv}. Our results suggest the use of the presented CLDNN and ResNet architectures at low SNR and the LSTM and ResNet at high SNR. In Section~\ref{sec:dimred}, we studied the problem of reducing the input dimensions for faster training. Our results suggest the use of PCA to reduce the input dimensions at low SNR, and subsampling techniques at high SNR. In particular, selecting the samples with the largest magnitude values leads to the highest classification accuracy at high SNR. It is worth noting here that it is straightforward to implement this magnitude rank subsampling for online training, by dynamically adjusting a threshold, and ignoring arriving samples whose magnitude values are below the threshold. Further, the problem of deriving an online version of PCA has been considered in the literature (see e.g.,~\cite{online-pca}). In Section~\ref{sec:snr}, we investigated the selection of representative SNR training values, while still testing the trained networks on the whole considered SNR range. The results support the effectiveness of training by SNR selection compared to a randomly selected training data set of the same size. For the considered range of SNR values from -20 dB to 18 dB, we found that choosing a pair of SNR values for training lead to superior classification accuracies over a wide range of testing SNR values.   
\subsection{Rectangular or Polar Form Representation}
We observe from the investigation in this work that we obtain better performance when the complex samples are represented in rectangular form, when input to all studied networks, except the pure LSTM, in which the input samples are better represented in polar form. We also observe that LSTM is particularly good - compared to other architectures - at distinguishing between different QAM constellations (see \cite{asilomar17} and \cite{lstm} for more details). The underlying reason is that these modulation types rely on small variations in amplitude and phase, and hence, using input data represented in polar form and an LSTM classifier that can identify repeating patterns of changes could deliver high classification accuracy for these modulation types. This sheds light on the sensitivity of neural network classifiers to the input representation. We plan to investigate for future work, whether one could alleviate the impact of this sensitivity by processing the data through two parallel architectures, and then adding one or more dense layers to learn which representation would lead to better classification performance for the task at hand.  

\subsection{Long-term Dependencies and Skip Connections}
Other than CNN, all the other neural networks presented in this work are obtained through modifications by either capturing long-term dependencies through LSTM layers , as in the CLDNN and pure LSTM architectures, or by adding shortcut connections between non-consecutive layers to mitigate the vanishing gradient problem and add flexibility to the architecture (see e.g., \cite{dl-book}), as in the DenseNet and ResNet architectures. We believe that capturing long-term dependencies is useful for the considered modulation recognition task because it helps identify repeating patterns of symbol-to-symbol transitions; such patterns could be used as a signature for the modulation type. Further, adding shortcut connections is also useful, because of the relatively large number of considered modulation types, and the varying nature of distinguishing between different pairs. While deeper networks could be useful for drawing distinct features between similar modulation types, this may cause overfitting for simpler tasks, which require shallower architectures that are obtained by activating the shortcut connections.

\subsection{Why Dimensionality Reduction leads to Fast Deep Learning}
Recent theoretical attempts to explain deep learning have found it particularly plausible to hypothesize that most of the training time is spent compressing the input data (see e.g., \cite{bottleneck}). We believe that this could be a key reason leading to our presented results in Section~\ref{sec:dimred} and Section~\ref{sec:snr}, particularly the minimal loss in accuracy that we found even with aggressive subsampling rates for the magnitude rank subsampling studied in Section~\ref{sec:rank}. For future work, we also plan to investigate using the hidden layer representation of autoencoders (see e.g., \cite[Chapter $14$]{dl-book}) for input data compression.

\subsection{Future Work: Denoising Autoencoders?}
We observe a strong similarity between the wireless communication problem, where uncertainty is introduced by unknown channel impairments, and the recently studied problem of defending neural networks against adversarial perturbations (see e.g., \cite{adv}). For example, an $l_2$ bounded attack would correspond to channel noise with power equivalent to the $l_2$ bound. For the latter problem, it was found recently that the use of denoising autoencoders is particularly effective~\cite{cascade-ae}. We plan to investigate using these autoencoder architectures for the considered problem. In particular, we expect them to allow us to obtain better classification accuracy at low SNR values, even when we the training data set only includes samples taken at high SNR values. 

\subsection{Minimalist Training Setup}
In Section~\ref{sec:dimred} and Section~\ref{sec:snr}, we demonstrated the possibility of significantly reducing the training time (by almost 20 times) while incurring minimal loss in accuracy (could be as low as 2\% at high SNR), through various ideas for minimizing the size of the training data set while preserving relevant information needed for learning. We believe that this opens the door for a line of research that aims at deploying deep learning algorithms for real-time autonomous wireless communications. For future work, we are interested in testing combinations of the presented ideas. For example, combining SNR selection using denoising autoencoder and combining sub-Nyquist sampling techniques with using  the hidden layer representation of deep autoencoders for dimensionality reduction. Based on the initial results presented in this work, we foresee great potential in such methods for making it feasible to train deep neural networks online for tasks essential to next-generation wireless communication systems.

\section{Concluding Remarks}\label{sec:conclusion}
In this work, we presented fast deep learning algorithms for distinguishing between 10 different modulation types, with high classification accuracy over a wide range of SNR values. We identified CLDNN and ResNet deep neural network architectures that perform best at low SNR, and LSTM and ResNet architectures that perform best at high SNR. Further, our results suggest the use of PCA to reduce input dimensions for faster training at low SNR, and subsampling based on sample magnitude values at high SNR. We finally identified representative training SNR values, and found that training with data sets corresponding to only two SNR values - one at high SNR and another at low SNR - in the range from -20 to 18 dB lead to achieving high classification accuracies over a wide portion of that range. For future work, our plan is to investigate the optimal combination of the presented algorithms, as well as new methods like using denoising autoencoders.
\bibliographystyle{IEEEtran}
\bibliography{refs.bib}

\begin{thebibliography}{10}
\providecommand{\url}[1]{#1}
\csname url@samestyle\endcsname
\providecommand{\newblock}{\relax}
\providecommand{\bibinfo}[2]{#2}
\providecommand{\BIBentrySTDinterwordspacing}{\spaceskip=0pt\relax}
\providecommand{\BIBentryALTinterwordstretchfactor}{4}
\providecommand{\BIBentryALTinterwordspacing}{\spaceskip=\fontdimen2\font plus
\BIBentryALTinterwordstretchfactor\fontdimen3\font minus
  \fontdimen4\font\relax}
\providecommand{\BIBforeignlanguage}[2]{{%
\expandafter\ifx\csname l@#1\endcsname\relax
\typeout{** WARNING: IEEEtran.bst: No hyphenation pattern has been}%
\typeout{** loaded for the language `#1'. Using the pattern for}%
\typeout{** the default language instead.}%
\else
\language=\csname l@#1\endcsname
\fi
#2}}
\providecommand{\BIBdecl}{\relax}
\BIBdecl

\bibitem{asilomar17}
X.~Liu, D.~Yang, and A.~{El Gamal}, ``Deep neural network architectures for
  modulation classification,'' in \emph{Proc. IEEE Asilomar Conference on
  Signals, Systems and Computers}, 2017.

\bibitem{sills1999maximum}
J.~Sills, ``Maximum-likelihood modulation classification for psk/qam,'' in
  \emph{Proc. IEEE Military Communications Conference (MILCOM)}, 1999.

\bibitem{polydoros1990detection}
A.~Polydoros and K.~Kim, ``On the detection and classification of quadrature
  digital modulations in broad-band noise,'' \emph{IEEE Transactions on
  Communications}, vol.~38, no.~8, pp. 1199--1211, 1990.

\bibitem{sapiano1996maximum}
P.~Sapiano and J.~Martin, ``Maximum likelihood {PSK} classifier,'' in
  \emph{Proc. IEEE Military Communications Conference (MILCOM)}, 1996.

\bibitem{beidas1998asynchronous}
B.~F. Beidas and C.~L. Weber, ``Asynchronous classification of {MFSK} signals
  using the higher order correlation domain,'' \emph{IEEE Transactions on
  communications}, vol.~46, no.~4, pp. 480--493, 1998.

\bibitem{panagiotou2000likelihood}
P.~Panagiotou, A.~Anastasopoulos, and A.~Polydoros, ``Likelihood ratio tests
  for modulation classification,'' in \emph{Proc. IEEE Military Communications
  Conference (MILCOM)}, 2000.

\bibitem{hong2002antenna}
L.~Hong and K.~Ho, ``Antenna array likelihood modulation classifier for {BPSK}
  and {QPSK} signals,'' in \emph{Proc. IEEE Military Communications Conference
  (MILCOM)}, 2002.

\bibitem{hsue1989automatic}
S.-Z. Hsue and S.~S. Soliman, ``Automatic modulation recognition of digitally
  modulated signals,'' in \emph{Proc. IEEE Military Communications Conference
  (MILCOM)}, 1989.

\bibitem{hong1999identification}
L.~Hong and K.~Ho, ``Identification of digital modulation types using the
  wavelet transform,'' in \emph{Proc. IEEE Military Communications Conference
  (MILCOM)}, 1999.

\bibitem{swami2000hierarchical}
A.~Swami and B.~M. Sadler, ``Hierarchical digital modulation classification
  using cumulants,'' \emph{IEEE Transactions on communications}, vol.~48,
  no.~3, pp. 416--429, 2000.

\bibitem{hatzichristos2001hierarchical}
G.~Hatzichristos and M.~P. Fargues, ``A hierarchical approach to the
  classification of digital modulation types in multipath environments,'' in
  \emph{Proc. IEEE Asilomar Conference on Signals, Systems, and Computers},
  2001.

\bibitem{soliman1992signal}
S.~S. Soliman and S.-Z. Hsue, ``Signal classification using statistical
  moments,'' \emph{IEEE Transactions on Communications}, vol.~40, no.~5, pp.
  908--916, 1992.

\bibitem{lichun2002comments}
L.~Lichun, ``Comments on signal classification using statistical moments,''
  \emph{IEEE Transactions on Communications}, vol.~50, no.~2, p. 195, 2002.

\bibitem{mingquan1998ar}
L.~Mingquan, X.~Xianci, and L.~Lemin, ``{AR} modeling-based features extraction
  of multiple signals for modulation recognition,'' in \emph{Proc. IEEE
  International Conference on Signal Processing}, 1998.

\bibitem{mobasseri2000digital}
B.~G. Mobasseri, ``Digital modulation classification using constellation
  shape,'' in \emph{Proc. IEEE International Conference on Signal Processing},
  2000.

\bibitem{mingquan1996cyclic}
L.~Mingquan, X.~Xianci, and L.~Leming, ``Cyclic spectral features based
  modulation recognition,'' in \emph{Proc. International Conference on
  Communication Technology (ICCT)}, 1996.

\bibitem{azzouz1996modulation}
E.~E. Azzouz and A.~K. Nandi, ``Modulation recognition using artificial neural
  networks,'' \emph{Signal Processing}, vol.~56, no.~2, pp. 165--175, 1997.

\bibitem{nolan2001modulation}
K.~E. Nolan, L.~Doyle, D.~O'Mahony, and P.~Mackenzie, ``Modulation scheme
  recognition techniques for software radio on a general purpose processor
  platform,'' in \emph{Proc. Joint IEI/IEE Symposium on Telecommunication
  Systems, Dublin}, 2001.

\bibitem{kim1988digital}
K.~Kim and A.~Polydoros, ``Digital modulation classification: the {BPSK} versus
  {QPSK} case,'' in \emph{Proc. IEEE Military Communications Conference
  (MILCOM)}, 1988.

\bibitem{lay1994per}
N.~E. Lay and A.~Polydoros, ``Per-survivor processing for channel acquisition,
  data detection and modulation classification,'' 1994.

\bibitem{park2008automatic}
C.-S. Park, J.-H. Choi, S.-P. Nah, W.~Jang, and D.~Y. Kim, ``Automatic
  modulation recognition of digital signals using wavelet features and {SVM},''
  in \emph{Proc. International Conference on Advanced Communications
  Technology}, 2008.

\bibitem{de2010prototype}
L.~De~Vito, S.~Rapuano, and M.~Villanacci, ``Prototype of an automatic digital
  modulation classifier embedded in a real-time spectrum analyzer,'' \emph{IEEE
  Transactions on Instrumentation and Measurement}, vol.~59, no.~10, pp.
  2639--2651, 2010.

\bibitem{conv}
T.~{O'Shea}, J.~Corgan, and T.~Clancy, ``Convolutional radio modulation
  recognition networks,'' in \emph{Proc. International conference on
  engineering applications of neural networks}, 2016.

\bibitem{resnet}
K.~He, X.~Zhang, S.~Ren, and J.~Sun, ``Deep residual learning for image
  recognition,'' in \emph{Proc. IEEE Conference on Computer Vision and Pattern
  Recognition (CVPR)}, 2016.

\bibitem{densenet}
G.~Huang, Z.~Liu, L.~Van Der~Maaten, and K.~Q. Weinberger, ``Densely connected
  convolutional networks.'' in \emph{Proc. IEEE Conference on Computer Vision
  and Pattern Recognition (CVPR)}, 2017.

\bibitem{new-resnet}
T.~{O'Shea}, T.~James, T.~Roy, and T.~Clancy, ``Over-the-air deep learning
  based radio signal classification,'' \emph{IEEE Journal of Selected Topics in
  Signal Processing}, vol.~12, no.~1, pp. 168--179, 2018.

\bibitem{CLDNN}
T.~N. Sainath, O.~Vinyals, A.~W. Senior, and H.~Sak, ``Convolutional, long
  short-term memory, fully connected deep neural networks,'' in \emph{Proc.
  IEEE International Conference on Acoustics, Speech and Signal Processing
  (ICASSP)}, 2015.

\bibitem{west2017deep}
N.~E. West and T.~O'Shea, ``Deep architectures for modulation recognition,'' in
  \emph{International Symposium on Dynamic Spectrum Access Networks (DySPAN)},
  2017.

\bibitem{datagen}
T.~O'Shea and N.~West, ``Radio machine learning dataset generation with gnu
  radio,'' in \emph{Proc. GNU Radio Conference}, 2016.

\bibitem{yonina-book}
Y.~C. Eldar, \emph{Sampling Theory: {Beyond} Bandlimited Systems}.\hskip 1em
  plus 0.5em minus 0.4em\relax Cambridge University Press, 2015.

\bibitem{dl-book}
I.~Goodfellow, Y.~Bengio, and A.~Courville, \emph{Deep learning}.\hskip 1em
  plus 0.5em minus 0.4em\relax MIT Press, 2016.

\bibitem{lstm}
S.~Rajendran, W.~Meert, D.~Giustiniano, V.~Lenders, and S.~Pollin, ``Deep
  learning models for wireless signal classification with distributed low-cost
  spectrum sensors,'' \emph{IEEE Transactions on Cognitive Communications and
  Networking}, vol.~4, no.~3, pp. 433--445, 2018.

\bibitem{pca}
K.~Pearson, ``On lines and planes of closest fit to systems of points in
  space,'' \emph{Philosophical Magazine}, vol.~2, no.~11, pp. 559--572, 1901.

\bibitem{online-pca}
C.~Boutsidis, D.~Garber, Z.~Karnin, and E.~Liberty, ``Online principal
  component analysis,'' Available at:
  http://cs-www.cs.yale.edu/homes/el327/papers/opca.pdf.

\bibitem{bottleneck}
A.~M. Saxe, Y.~Bansal, J.~Dapello, M.~Advani, A.~Kolchinsky, B.~Tracey, and
  D.~Cox, ``On the information bottleneck theory of deep learning,'' in
  \emph{Proc. International Conference on Learning Representations (ICLR)},
  2018.

\bibitem{adv}
L.~Huang, A.~D. Joseph, B.~Nelson, B.~Rubinstein, and J.~Tygar, ``Adversarial
  machine learning,'' in \emph{Proc. AMC Workshop on Security and Artificial
  Intelligence}, 2011.

\bibitem{cascade-ae}
R.~Sahay, R.~Mahfuz, and A.~{El Gamal}, ``Combatting adversarial attacks
  through denoising and dimensionality reduction: {A} cascaded autoencoder
  approach,'' Arxiv preprint arXiv:1812.03087, Dec. 2018.

\end{thebibliography}
\end{document}